\documentclass[lettersize,journal]{IEEEtran}
\usepackage{amsmath,amsfonts,amssymb}
\usepackage{algorithmic}
\usepackage{algorithm}
\usepackage{array}
\usepackage[caption=false,font=footnotesize,labelfont=rm,textfont=rm]{subfig}
\usepackage{textcomp}
\usepackage{stfloats}
\usepackage{url}
\usepackage{verbatim}
\usepackage{graphicx}
\usepackage{cite}
\usepackage{color}
\begin{document}

\title{Movable Antenna-Aided Secure Full-Duplex Multi-User Communications}

\author{Jingze Ding, \IEEEmembership{Graduate Student Member, IEEE},
		Zijian Zhou, \IEEEmembership{Member, IEEE}, \\
		and Bingli Jiao, \IEEEmembership{Senior Member, IEEE}
\thanks{Part of this work has been accepted by IEEE GLOBECOM 2024\cite{Ding2}.}
\thanks{This work was supported by National Natural Science Foundation of China under Grant 62171006.  The calculations were supported by High-Performance Computing Platform of Peking University. \textit{(Corresponding authors: Jingze Ding; Zijian Zhou.)}}
\thanks{Jingze Ding and Bingli Jiao are with the School of Electronics, Peking University, Beijing 100871, China (e-mail: djz@stu.pku.edu.cn; jiaobl@pku.edu.cn).}
\thanks{Zijian Zhou is with the School of Science and Engineering, The Chinese University of Hong Kong, Shenzhen, Guangdong 518172, China (e-mail: zjzhou@cuhk.edu.cn).}
}

\maketitle

\begin{abstract}
In this paper, we investigate physical layer security (PLS) for full-duplex (FD) multi-user systems. We consider a base station (BS) that operates in FD mode and transmits artificial noise (AN) to simultaneously protect uplink (UL) and downlink (DL) transmissions. Conventional fixed-position antennas (FPAs) at the FD BS struggle to fully exploit spatial degrees of freedom (DoFs) to improve signal reception and suppress interference. To overcome this limitation, we propose a novel FD BS architecture equipped with multiple transmit and receive movable antennas (MAs). The MAs introduce the DoFs in antenna position optimization, which can improve the performance of secure communication systems. To serve users and counter the cooperative interception of multiple eavesdroppers (Eves), we formulate a sum of secrecy rates (SSR) maximization problem to jointly optimize the MA positions, the transmit, receive, and AN beamformers at the BS, and the UL powers. We propose an alternating optimization (AO) algorithm, which decomposes the original problem into three sub-problems, to solve the challenging non-convex optimization problem with highly coupled variables. Specifically, we propose the multi-velocity particle swarm optimization (MVPSO), which is an improved version of the standard particle swarm optimization (PSO), to simultaneously optimize all MA positions. The transmit/AN beamformers and the UL powers are solved by successive convex approximation (SCA). The optimal receive beamformer is derived as a closed-form solution. Simulation results demonstrate the effectiveness of the proposed algorithms and the advantages of MAs over conventional FPAs in enhancing the security of FD multi-user systems.
\end{abstract}


\section{Introduction}
\IEEEPARstart{I}{n} the era of ubiquitous connectivity, people extensively rely on wireless networks to transmit important and private information. Consequently, security and privacy have become critical concerns for next-generation wireless communications \cite{survey1}. Over the last few decades, conventional security mechanisms primarily focus on cryptographic encryption/decryption methods \cite{key}. However, as threats evolve and become increasingly sophisticated, the encryption/decryption-based methods will cause heavy computation and key management costs. Thus, relying merely on these solutions is insufficient. This has led to a growing interest in physical layer security (PLS), which is based on information theory fundamentals and emphasizes the secrecy capacity of propagation channels \cite{survey2,survey3}.

In particular, a pivotal technique for enhancing PLS is multi-antenna transmission, which leverages spatial degrees of freedom (DoFs) \cite{survey4}. To improve legitimate channels while degrading eavesdropping channels, significant efforts have been devoted to secure beamforming techniques in various scenarios \cite{multiantenna1,multiantenna2,multiantenna3}. By exploiting the spatial diversity offered by multiple antennas, these studies comprehensively validated the efficacy of beamforming in improving security performances.

However, conventional multi-antenna systems typically utilize fixed-position antennas (FPAs), which restrict their abilities to further exploit the channel variations, especially in cases with a limited number of antennas. To overcome this limitation, movable antenna (MA) \cite{MA1} offers a practical and innovative solution. The MA enables flexible movement within two or three-dimensional region through a driver, such as a step motor along a slide track \cite{MA2}. Due to continuous movement, the MA can better utilize spatial DoFs than the antenna selection (AS) \cite{AS} with discretely arranged antennas.  So far, numerous studies have demonstrated the advantages of MA-aided systems over FPA systems. The pioneering work \cite{MA1} established the field-response channel model for MA-aided communication systems. Under the far-field condition, the authors provided a methodology for calculating the channel responses at various MA positions and analyzed the variations in channel gains under deterministic and statistical channels. Moreover, the authors in \cite{MA3} investigated the channel capacity of multiple-input multiple-output (MIMO) systems with MAs. Dealing with the challenges in multi-user uplink (UL) communications, the authors in \cite{MA4,MA5} respectively considered the utilization of MAs at the base station (BS) and user ends. Regarding multi-user downlink (DL) communications, the authors in \cite{MA6} modeled the motions of MAs as discrete movements and jointly optimized the transmit beamformer and the MA positions at the BS to minimize the total transmit power. Furthermore, the authors in \cite{MA7,MA8} investigated the MA array-aided beamforming. The antenna position and weight vectors are jointly optimized to achieve the full array gain and the null steering over the desired and undesired directions, respectively. Moreover, the MA-aided secure communication system has become a hot topic. Specifically, based on the obtained channel state information (CSI) and the previously proposed field-response channel model, the channel responses at any location in a given region, i.e., channel gain map, can be reconstructed. Then, based on the channel gain map, the MAs can move to positions where the channel power gains are advantageous/high for legitimate users but adverse/low for eavesdroppers (Eves), leading to an improved secrecy rate. Reference \cite{MA9} showed the performance improvements in multiple-input single-output (MISO) systems achieved by the one-dimensional MA array over conventional FPAs. The authors in \cite{MA10} extended the movements of MAs to two dimensions and studied the secure DL communication in a MISO system with a single-FPA user and a single-FPA Eve. Furthermore, a more general MA-aided secure MIMO DL communication system was investigated in \cite{MA11}. Besides, without perfect CSI of the Eves, the authors in \cite{MA12,MA13} jointly optimized the transmit beamformer and MA positions to bolster systems' securities. It is worth noting that accurate CSI is paramount to ensuring the improvements of MA-aided communication systems. Fortunately, recent works \cite{MA14,MA15} have proposed some practical methods with low pilot overhead and computational complexity to achieve satisfactory CSI estimation for MA-aided systems.

In addition, artificial noise (AN) is another effective technique to improve the secrecy rate by interfering with Eves' receptions \cite{AN1}. Nevertheless, in half-duplex (HD) mode, the AN transmitted by the BS solely shields the DL users, leaving the UL users entirely vulnerable to the Eves, particularly when each UL user is equipped with only a single FPA. It is noteworthy that the full-duplex (FD) mode, wherein transmit and receive signals are superimposed onto the same time-frequency resource block, naturally addresses this issue while concurrently yielding gains in spectral efficiency \cite{light}. Consequently, the resource allocations for secure FD multi-user systems employing AN have garnered increasing attention. Reference \cite{AN2} proposed the robust beamforming and jamming methods using AN in a worst case, where an FD BS simultaneously serves multiple UL and DL users with single FPA in the presence of a multi-FPA Eve. The authors in \cite{AN3} expanded the scenario to include multiple Eves with single FPA. Additionally, the authors in \cite{AN4} considered a more complex scenario involving multiple UL users, DL users, and multi-FPA Eves and jointly minimized the total transmit powers for achieving secure UL and DL transmissions concurrently.

As previously mentioned, current research on MA-aided secure communication systems is confined to HD mode, which fails to provide comprehensive protection for both UL and DL users. On the other hand, existing secure FD systems have solely considered FPAs, which limit the exploitation of spatial DoFs, and their results are not directly applicable to MA-aided systems with reconfigurable channels by antenna position optimization. We note that the mobility of MA provides remarkable advantages over conventional FPA in signal power enhancement, interference mitigation, and flexible beamforming \cite{MA2,Ding3}. Compared to HD mode, FD systems introduce additional self-interference (SI) with much higher power than the signal of interest (SoI) \cite{Ding1}. Therefore, the MAs are supposed to relocate to the positions where the SI channel gain is minimized and the SoI channel gain is maximized to suppress the SI and improve the power of SoI. Meanwhile, the geometry of an MA array can be reshaped to achieve multi-beamforming with high array gains over the target directions and low array gains over the undesired directions \cite{MA8}, which can help to mitigate the multi-user interference and AN interference, thereby leading to enhanced security performance. Currently, there is still a lack of system design for secure FD multi-user systems aided by MAs. As such, in this paper, we present the first exploration of PLS for MA-aided FD systems. Compared to existing HD MA-aided multi-access systems, e.g., \cite{MA5}, the FD system can not only enhance spectral efficiency but also protect both UL and DL transmissions. Thus, we aim to leverage the reconfigurability of MA positions to further improve the security of FD multi-user systems. The main contributions of this paper are summarized as follows.
\begin{itemize}
	\item [1)] 
	We propose a novel FD BS architecture equipped with separate transmit and receive MAs to simultaneously serve multiple UL and DL users while actively resisting cooperative eavesdropping by multiple Eves. An optimization problem is formulated to maximize the sum of secrecy rates (SSR) by jointly optimizing the MA positions, the transmit, receive, and AN beamformers at the BS, and the UL powers, subject to the constraints of maximum transmit powers of each user and the BS, finite moving regions for MAs, and minimum inter-MA distance.
	\item [2)]
	We propose an alternating optimization (AO) algorithm to solve the formulated non-convex optimization problem with highly coupled variables. The AO decomposes the original problem into three sub-problems and iteratively solves them. In particular, we modify the standard particle swarm optimization (PSO) and propose the multi-velocity PSO (MVPSO) for optimizing MA positions to avoid undesired sub-optimal solutions. Then, we reformulate the SSR into a more tractable form and utilize the successive convex approximation (SCA) to optimize the transmit and AN beamformers at the BS and the UL powers. Finally, the closed-form expression for the optimal receive beamformer is derived.
	\item [3)]
	We conduct extensive simulations to evaluate the effectiveness of the proposed algorithm and the advantages of the proposed scheme for the MA-aided secure FD multi-user system. Simulation results confirm that the proposed MVPSO algorithm effectively avoids undesired local optimal solutions compared to the standard PSO algorithm and generally outperforms the state-of-the-art antenna position optimization algorithms. Besides, compared to the conventional HD BS with only FPAs, the proposed FD BS equipped with MAs can excellently utilize spatial DoFs to resist the Eves and assist the UL and DL communications. Meanwhile, the reuse of time-frequency resources in FD mode and the AN released by the BS further improve the SSR. Moreover, the derived optimal receive beamformer significantly strengthens the reception of UL signals. Finally, we evaluate the impact of the imperfect field-response information (FRI) on MA positioning.
\end{itemize}

The rest of this paper is organized as follows. Section \ref{2} introduces the system model and the optimization problem for the proposed system. In Section \ref{3}, we propose the AO algorithm to solve the optimization problem. Next, simulation results and discussions are provided in Section \ref{4}. Finally, this paper is concluded in Section \ref{5}.

\textit{Notation:}  $a/A$, $\mathbf{a}$, $\mathbf{A}$, and $\mathcal{A}$ denote a scalar, a vector, a matrix, and a set, respectively. $ \left[\mathbf{a}\right]_i$ denotes the $i$-th element of vector $\mathbf{a}$. $\mathbf{A} \succeq 0$ indicates that $\mathbf{A}$ is a positive semidefinite matrix. ${\left(  \cdot  \right)^{T}}$, ${\left(  \cdot  \right)^{H}}$, $\left|  \cdot  \right|$, $\left\|  \cdot  \right\|_2$, $\mathrm{Tr}\left\lbrace  \cdot\right\rbrace  $, and $\mathrm{Rank}\left\lbrace  \cdot\right\rbrace  $ denote the transpose, conjugate transpose, absolute value, Euclidean norm, trace, and rank, respectively. $\odot$ represents Hadamard product. $\left[ x \right]^+$ stands for $\max\left\lbrace 0, x\right\rbrace $. $\mathbb{C}^{M \times N}$ and $\mathbb{R}^{M \times N}$ are the sets for complex and real matrices of $M \times N$ dimensions, respectively. $\mathbf{I}_N$ is the identity matrix of order $N$. $\mathcal{CN}\left( \mathbf{0}, \mathbf{\Lambda} \right) $ represents the circularly symmetric complex Gaussian (CSCG) distribution with mean zero and covariance matrix $\mathbf{\Lambda}$. $\mathcal{A} \backslash \mathcal{B}$ denotes the subtraction of set $\mathcal{B}$ from set $\mathcal{A}$. $\sim$ and $\triangleq$ stand for ``distributed as'' and ``defined as'', respectively.
\section{System Model}\label{2}
\begin{figure*}[!t]
	\centering
	\includegraphics[width=2\columnwidth]{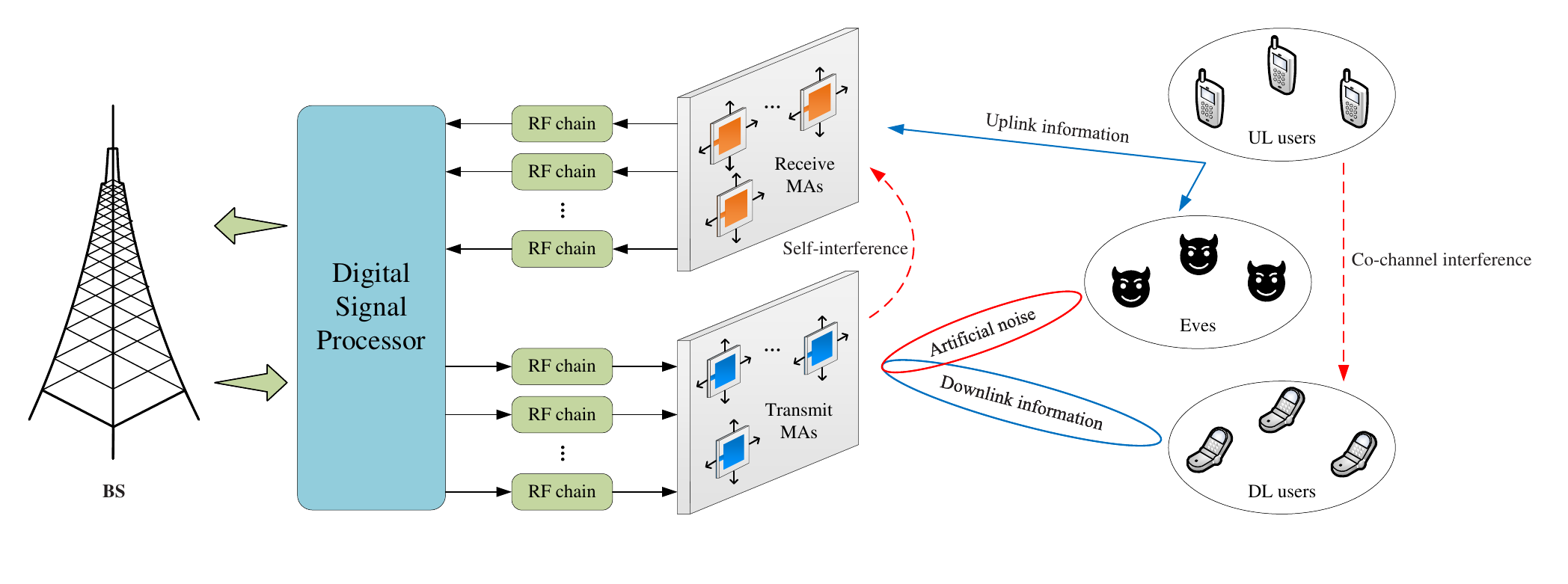}
	\caption{Illustration of the proposed MA-aided secure FD multi-user system.}
	\label{system_model}
\end{figure*}
\begin{figure}[!t]
	\centering
	\includegraphics[width=1\columnwidth]{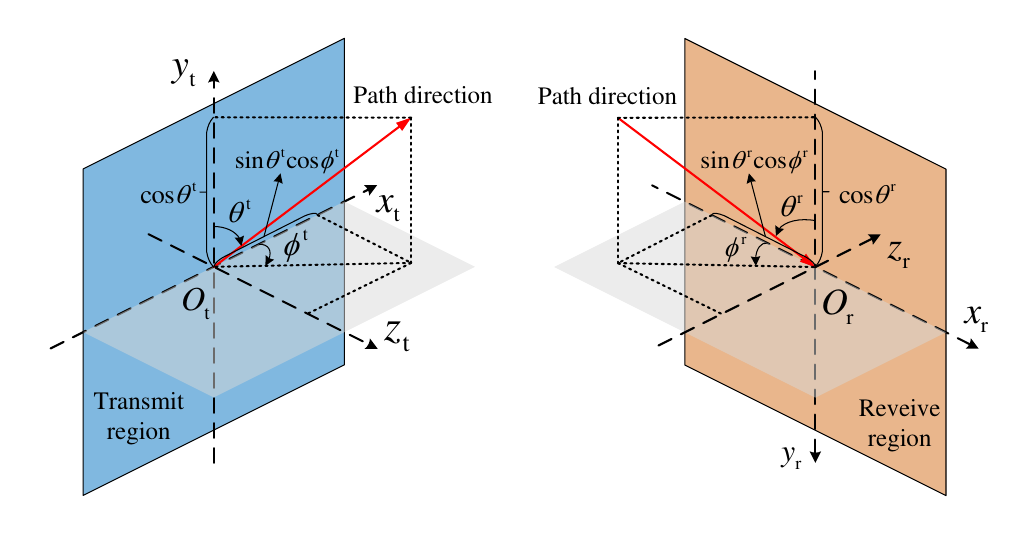}
	\caption{Illustration of the coordinates and spatial angles for transmit and receive regions.}
	\label{channel_model}
\end{figure}
As shown in Fig.\;\ref{system_model}, we consider a secure FD multi-user system with an MA-aided FD BS to serve $K_\mathrm{U}$ HD UL users and $K_\mathrm{D}$ HD DL users, in the presence of $K_\mathrm{E}$ HD Eves. Each user or Eve is equipped with a single FPA. The BS is equipped with $N_\mathrm{t}$ transmit MAs and $N_\mathrm{r}$ receive MAs, which can move in the two-dimensional regions for actively reconfiguring the channel conditions\footnote{We will consider a quasi-static block-fading channel model with static or low-mobility terminals in Section \ref{channel}. Thus, the power consumption for antenna movement is negligible due to infrequent MA position changes. For the scenarios with fast-fading channels, the MA positions can be designed based on the statistical CSI to avoid frequent antenna movements.}. The positions of the ${n_\mathrm{t}}$-th transmit MA and the ${n_\mathrm{r}}$-th receive MA are described by their Cartesian coordinates, i.e., ${\mathbf{t}_{{n_\mathrm{t}}}} = {\left[ {{x^{\mathrm{t}}_{n_\mathrm{t}}},{y^{\mathrm{t}}_{n_\mathrm{t}}}} \right]^T} \in {\mathcal{C}_\mathrm{t}}$ ($1 \le {n_\mathrm{t}} \le {N_\mathrm{t}}$) and ${\mathbf{r}_{{n_\mathrm{r}}}} = {\left[ {{x^{\mathrm{r}}_{n_\mathrm{r}}},{y^{\mathrm{r}}_{n_\mathrm{r}}}} \right]^T} \in {\mathcal{C}_\mathrm{r}}$ ($1 \le {n_\mathrm{r}} \le {N_\mathrm{r}}$), where ${\mathcal{C}_\mathrm{t}}$ and ${\mathcal{C}_\mathrm{r}}$ represent the transmit and receive regions, respectively.

Define the collections of $N_\mathrm{t}$ transmit MAs and $N_\mathrm{r}$ receive MAs as $\tilde {\mathbf{t}} = {\left[ {\mathbf{t}_1^T, \cdots ,\mathbf{t}_{{N_\mathrm{t}}}^T} \right]^T} \in {\mathbb{R}^{2{N_\mathrm{t}} \times 1}}$ and $\tilde {\mathbf{r}} = {\left[ {\mathbf{r}_1^T, \cdots ,\mathbf{r}_{{N_\mathrm{r}}}^T} \right]^T} \in {\mathbb{R}^{2{N_\mathrm{r}} \times 1}}$, respectively. For MA-aided communication systems, the channel response can be written as the function of MA positions \cite{MA1}. Thus, the SI channel of the BS, the channel from the BS to DL user $k_\mathrm{D}$ ($k_\mathrm{D} \in \left\{ {1, \ldots ,K_\mathrm{D}} \right\} \triangleq  \mathcal{K}_\mathrm{D}$), the channel from the BS to Eve $k_\mathrm{E}$ ($k_\mathrm{E} \in \left\{ {1, \ldots ,K_\mathrm{E}} \right\} \triangleq  \mathcal{K}_\mathrm{E}$), and the channel from UL user $k_\mathrm{U}$ ($k_\mathrm{U} \in \left\{ {1, \ldots ,K_\mathrm{U}} \right\} \triangleq  \mathcal{K}_\mathrm{U}$) to the BS are denoted as ${\mathbf{H}_\mathrm{SI}} \left( {\tilde{ \mathbf{t}}}, {\tilde{ \mathbf{r}}} \right) \in {\mathbb{C}^{{N_\mathrm{r}} \times {N_\mathrm{t}}}}$, $\mathbf{h}_{\mathrm{BD},{k_\mathrm{D}}}\left( \tilde {\mathbf{t}} \right) \in {\mathbb{C}^{{N_\mathrm{t}} \times 1}}$, $\mathbf{h}_{\mathrm{BE},{k_\mathrm{E}}} \left( \tilde {\mathbf{t}}\right) \in {\mathbb{C}^{{N_\mathrm{t}} \times 1}}$, and ${\mathbf{h}_{\mathrm{UB},{k_\mathrm{U}}}\left( \tilde {\mathbf{r}} \right) \in {\mathbb{C}^{{N_\mathrm{r}} \times 1}}}$, respectively.  Besides, we denote the channel from UL user ${k_\mathrm{U}}$ to DL user ${k_\mathrm{D}}$ as ${h_{\mathrm{UD},{k_\mathrm{U}},{k_\mathrm{D}}}} \in {\mathbb{C}^{1 \times 1}}$ and to Eve ${k_\mathrm{E}}$ as ${h_{\mathrm{UE},{k_\mathrm{U}},{k_\mathrm{E}}}} \in {\mathbb{C}^{1 \times 1}}$.

In a given time slot, the BS transmits ${K_\mathrm{D}}$ independent information streams to the ${K_\mathrm{D}}$ DL users. For DL user ${k_\mathrm{D}}$, the information stream can be expressed as ${\mathbf{x}_{k_\mathrm{D}}} = {\mathbf{w}_{k_\mathrm{D}}}{s_{k_\mathrm{D}}^{\mathrm{D}}}$, where ${s_{k_\mathrm{D}}^{\mathrm{D}}}$ denote the DL information with normalized power, and ${\mathbf{w}_{k_\mathrm{D}}} \in {\mathbb{C}^{{N_\mathrm{t}} \times 1}}$ is the corresponding beamformer\footnote{In DL multi-user communication scenarios, each user may have independent data requirements, $s_{k_\mathrm{D}}^{\mathrm{D}}$. Thus, the BS needs to transmit all users' data simultaneously and use the corresponding beamformers, $\mathbf{w}_{k_\mathrm{D}}$, to steer the signals toward the intended receive terminals.}. To ensure security, the BS also generates an AN vector ${\mathbf{v}} \in {\mathbb{C}^{{N_\mathrm{t}} \times 1}}$ and transmits it with DL information to interfere with Eves' malicious interception\footnote{We will optimize the AN beamformer, $\mathbf{v}$, in Section \ref{opt_wv} to achieve high array gains over Eves' directions and low array gains over users' directions, thereby reducing the eavesdropping signal-to-interference-plus-noise ratios (SINRs) in \eqref{SINR_EU} and \eqref{SINR_ED}.}. Therefore, the DL signals can be expressed as
\begin{equation}
	\mathbf{x} = \sum\limits_{{k_\mathrm{D}} \in {\mathcal{K}_\mathrm{D}}} {{\mathbf{x}_{k_\mathrm{D}}}}  + \mathbf{v} \in {\mathbb{C}^{{N_\mathrm{t}} \times 1}}.
\end{equation}
\begin{figure*}[!b]
	\textsc{\centering
		\hrulefill
		\begin{align}
			& {\mathbf{y}^\mathrm{U}} = \underbrace {\sum\limits_{{k_\mathrm{U}} \in \mathcal{K}_\mathrm{U}} {{\mathbf{h}_{\mathrm{UB},{k_\mathrm{U}}}}\left( \tilde {\mathbf{r}} \right)\sqrt {{p_{k_\mathrm{U}}}} {s_{k_\mathrm{U}}^{\mathrm{U}}}} }_\mathrm{UL \ information} + \underbrace {{\mathbf{H}_\mathrm{SI}}\left( \tilde {\mathbf{t}}, \tilde {\mathbf{r}} \right)\sum\limits_{{k_\mathrm{D}} \in \mathcal{K}_\mathrm{D}} {{\mathbf{x}_{k_\mathrm{D}}}} + {\mathbf{H}_\mathrm{SI}}\left( \tilde {\mathbf{t}}, \tilde {\mathbf{r}} \right) \mathbf{v}}_\mathrm{Self-interference} + {\mathbf{n}^\mathrm{U}}. \label{BS} \\
			& {y_{k_\mathrm{D}}^{\mathrm{D}}} = \underbrace {\mathbf{h}_{\mathrm{BD},{k_\mathrm{D}}}^H \left( \tilde {\mathbf{t}} \right){\mathbf{x}_{k_\mathrm{D}}}}_\mathrm{DL \ desired \ information} + \underbrace {\sum\limits_{i \in {\mathcal{K}_\mathrm{D} \backslash \left\lbrace {k_\mathrm{D}}\right\rbrace }} {\mathbf{h}_{\mathrm{BD},{k_\mathrm{D}}}^H\left( \tilde {\mathbf{t}} \right){\mathbf{x}_i}} }_\mathrm{DL \ multi-user \ interference} + \underbrace {\mathbf{h}_{\mathrm{BD},{k_\mathrm{D}}}^H \left( \tilde {\mathbf{t}} \right) \mathbf{v}}_\mathrm{AN \ interference} + \underbrace {\sum\limits_{{k_\mathrm{U}} \in {\mathcal{K}_\mathrm{U}}} {h_{\mathrm{UD},{k_\mathrm{U}},{k_\mathrm{D}}}\sqrt {{p_{k_\mathrm{U}}}} } {s_{k_\mathrm{U}}^{\mathrm{U}}}}_\mathrm{Co-channel \ interference} + {n_{k_\mathrm{D}}^{\mathrm{D}}}. \label{DL} \\
			& {y_{k_\mathrm{E}}^{\mathrm{E}}} = \underbrace {\sum\limits_{{k_\mathrm{D}} \in \mathcal{K}_\mathrm{D}} {\mathbf{h}_{\mathrm{BE},{k_\mathrm{E}}}^H \left( \tilde {\mathbf{t}}\right){\mathbf{x}_{k_\mathrm{D}}}} }_\mathrm{DL \ information} + \underbrace {\mathbf{h}_{\mathrm{BE},{k_\mathrm{E}}}^H \left( \tilde {\mathbf{t}}\right) \mathbf{v} }_\mathrm{AN} + \underbrace {\sum\limits_{{k_\mathrm{U}} \in \mathcal{K}_\mathrm{U}} {{h_{\mathrm{UE},{k_\mathrm{U}},{k_\mathrm{E}}}}\sqrt {{p_{k_\mathrm{U}}}} {s_{k_\mathrm{U}}^{\mathrm{U}}}} }_\mathrm{UL \ information}  + {n_{k_\mathrm{E}}^{\mathrm{E}}} . \label{Eve}
	\end{align}}
\end{figure*}Then, the receive signals at the BS, DL user ${k_\mathrm{D}}$, and Eve ${k_\mathrm{E}}$ are respectively given by \eqref{BS}, \eqref{DL}, and \eqref{Eve} at the bottom of the page, where ${s_{k_\mathrm{U}}^{\mathrm{U}}}$ and ${p_{k_\mathrm{U}}}$ are the UL information with normalized power and the transmit power of UL user ${k_\mathrm{U}}$, respectively. ${\mathbf{n}^\mathrm{U}} \sim \mathcal{CN}\left( {\mathbf{0},\sigma _\mathrm{U}^2{\mathbf{I}_{{N_\mathrm{r}}}}} \right)$, ${n_{k_\mathrm{D}}^{\mathrm{D}}} \sim \mathcal{CN}\left( {0,\sigma _{\mathrm{D},{k_\mathrm{D}}}^2} \right)$, and ${n_{k_\mathrm{E}}^{\mathrm{E}}} \sim \mathcal{CN}\left( {0,\sigma _{\mathrm{E},{k_\mathrm{E}}}^2} \right)$ represent the additive white Gaussian
noise (AWGN) at the BS, DL user ${k_\mathrm{D}}$, and Eve ${k_\mathrm{E}}$ with average noise powers $\sigma _\mathrm{U}^2$, $\sigma _{\mathrm{D},{k_\mathrm{D}}}^2$, and $\sigma _{\mathrm{E},{k_\mathrm{E}}}^2$, respectively. 
\subsection{Channel Model}\label{channel}
We consider quasi-static block-fading channels and concentrate on one specific fading block with the multi-path channel components at any location in the regions given as fixed \cite{MA10}. As shown in Fig.\;\ref{channel_model}, define the elevation and azimuth angles of departure (AoDs) and angles of arrival (AoAs) as ${\theta ^\mathrm{t},\phi ^\mathrm{t}} \in \left[ {0,\pi } \right]$ and ${\theta ^\mathrm{r},\phi ^\mathrm{r}} \in \left[ {0,\pi } \right]$, respectively. Based on the field-response channel model \cite{MA1}, we establish the channel responses involving the MA, i.e., SI channel ${\mathbf{H}_\mathrm{SI}} \left( {\tilde{ \mathbf{t}}}, {\tilde{ \mathbf{r}}} \right)$, UL channel ${\mathbf{h}_{\mathrm{UB},k_\mathrm{U}}} \left( {\tilde{ \mathbf{r}}} \right)$, and DL channels ${\mathbf{h}_{\mathrm{BD},k_\mathrm{D}}} \left( {\tilde{ \mathbf{t}}} \right)$ and ${\mathbf{h}_{\mathrm{BE},k_\mathrm{E}}} \left( {\tilde{ \mathbf{t}}} \right)$, as follows.
\subsubsection{SI channel}
Let ${{L}_\mathrm{SI}^\mathrm{t}}$ and ${{L}_\mathrm{SI}^\mathrm{r}}$ denote the numbers of transmit and receive paths, respectively. The difference of the signal propagation distance for the $l^\mathrm{t}$-th ($1 \le l^\mathrm{t} \le {{L}_\mathrm{SI}^\mathrm{t}}$) transmit path between the MA position ${\mathbf{t}_{{n_\mathrm{t}}}}$ and the origin of the transmit region, i.e., $O_\mathrm{t}$ in Fig.\;\ref{channel_model}, can be expressed as $\rho _{\mathrm{SI},l^\mathrm{t}}^\mathrm{t}\left( {{\mathbf{t}_{{n_\mathrm{t}}}}} \right) = x_{{n_\mathrm{t}}}^\mathrm{t}\sin \theta _{\mathrm{SI},l^\mathrm{t}}^\mathrm{t}\cos \phi _{\mathrm{SI},l^\mathrm{t}}^\mathrm{t} + y_{{n_\mathrm{t}}}^\mathrm{t}\cos \theta _{\mathrm{SI},l^\mathrm{t}}^\mathrm{t}$. Denoting $\lambda$ as the carrier wavelength, the phase difference is calculated by $\frac{{2\pi }}{\lambda } \rho _{\mathrm{SI},l^\mathrm{t}}^\mathrm{t}\left( {{\mathbf{t}_{{n_\mathrm{t}}}}} \right)$. Thus, the transmit field-response vector (FRV), which characterizes the phase differences of ${{L}_\mathrm{SI}^\mathrm{t}}$ transmit paths, are obtained as
\begin{equation}
	{\mathbf{g}_\mathrm{SI}}\left( {{\mathbf{t}_{{n_\mathrm{t}}}}} \right) = {\left[ {{e^{\mathrm{j}\frac{{2\pi }}{\lambda }\rho _{\mathrm{SI},1}^\mathrm{t}\left( {{\mathbf{t}_{{n_\mathrm{t}}}}} \right)}}, \ldots ,{e^{\mathrm{j} \frac{{2\pi }}{\lambda }\rho _{\mathrm{SI},L_\mathrm{SI}^\mathrm{t}}^\mathrm{t}\left( {{\mathbf{t}_{{n_\mathrm{t}}}}} \right)}}} \right]^T} \in {\mathbb{C}^{{L_\mathrm{SI}^\mathrm{t}} \times 1}}.
\end{equation} 

Similarly, the receive FRV is given by
\begin{equation}
	{\mathbf{f}_\mathrm{SI}}\left( {{\mathbf{r}_{{n_\mathrm{r}}}}} \right) = {\left[ {{e^{\mathrm{j}\frac{{2\pi }}{\lambda }\rho _{\mathrm{SI},1}^\mathrm{r}\left( {{\mathbf{r}_{{n_\mathrm{r}}}}} \right)}}, \ldots ,{e^{\mathrm{j} \frac{{2\pi }}{\lambda }\rho _{\mathrm{SI},L_\mathrm{SI}^\mathrm{r}}^\mathrm{r}\left( {{\mathbf{r}_{{n_\mathrm{r}}}}} \right)}}} \right]^T} \in {\mathbb{C}^{{L_\mathrm{SI}^\mathrm{r}} \times 1}},
\end{equation}
which represents the phase differences of ${{L}_\mathrm{SI}^\mathrm{r}}$ receive paths, where $\rho _{\mathrm{SI},l^\mathrm{r}}^\mathrm{r}\left( {{\mathbf{r}_{{n_\mathrm{r}}}}} \right) = x_{{n_\mathrm{r}}}^\mathrm{r}\sin \theta _{\mathrm{SI},l^\mathrm{r}}^\mathrm{r}\cos \phi _{\mathrm{SI},l^\mathrm{r}}^\mathrm{r} + y_{{n_\mathrm{r}}}^\mathrm{r}\cos \theta _{\mathrm{SI},l^\mathrm{r}}^\mathrm{r}$ ($1 \le l^\mathrm{r} \le {{L}_\mathrm{SI}^\mathrm{r}}$) is the difference of the signal propagation distance for the $l^\mathrm{r}$-th receive path between the MA position ${\mathbf{r}_{{n_\mathrm{r}}}}$ and the origin of the receive region, i.e., $O_\mathrm{r}$ in Fig.\;\ref{channel_model}.

Moreover, define the path-response matrix (PRM) as $\mathbf{\Sigma} \in {\mathbb{C}^{{{L}_\mathrm{SI}^\mathrm{r}} \times{{L}_\mathrm{SI}^\mathrm{t}}}}$, where the entry in the $l^\mathrm{r}$-th row and $l^\mathrm{t}$-th column represents the channel response between the $l^\mathrm{t}$-th transmit path and the $l^\mathrm{r}$-th receive path from $O_\mathrm{t}$ to $O_\mathrm{r}$. As a result, the SI channel matrix is obtained as
\begin{equation} \label{H_SI}
	{\mathbf{H}_\mathrm{SI}} \left( {\tilde{ \mathbf{t}}}, {\tilde{ \mathbf{r}}} \right) ={\mathbf{F}_\mathrm{SI}}{\left( {\tilde {\mathbf{r}}} \right)^H}\mathbf{\Sigma} {\mathbf{G}_\mathrm{SI}}\left( {\tilde {\mathbf{t}}} \right) ,
\end{equation}
where $\mathbf{F}_\mathrm{SI}\left( {\tilde{\mathbf{r}}} \right) = \left[ \mathbf{f}_\mathrm{SI}\left( {\mathbf{r}_1} \right), \cdots, \mathbf{f}_\mathrm{SI}\left( {\mathbf{r}_{N_\mathrm{r}}} \right) \right] \in {\mathbb{C}^{{L_\mathrm{SI}^\mathrm{r}} \times N_\mathrm{r}}}$ and $\mathbf{G}_\mathrm{SI}\left( {\tilde{\mathbf{t}}} \right) = \left[ \mathbf{g}_\mathrm{SI}\left( {\mathbf{t}_1} \right), \cdots, \mathbf{g}_\mathrm{SI}\left( {\mathbf{t}_{N_\mathrm{t}}} \right) \right] \in {\mathbb{C}^{{L_\mathrm{SI}^\mathrm{t}} \times N_\mathrm{t}}}$ are the field-response matrices (FRMs) of $N_\mathrm{t}$ transmit MAs and $N_\mathrm{r}$ receive MAs, respectively.
\subsubsection{UL and DL Channels}
Since the users and Eves are equipped with FPAs, the FRMs only present at the receive end for UL channel and the transmit end for DL channel.
Denote $L_{\mathrm{UB},k_\mathrm{U}}^\mathrm{r}$, $L_{\mathrm{BD},k_\mathrm{D}}^\mathrm{t}$, and $L_{\mathrm{BE},k_\mathrm{E}}^\mathrm{t}$ as the numbers of the receive paths from UL user $k_\mathrm{U}$ to the BS and the transmit paths from the BS to DL user $k_\mathrm{D}$ and to Eve $k_\mathrm{E}$, respectively. Then, the UL and DL channels can be written as
\begin{align}
	{\mathbf{h}_{\mathrm{UB},k_\mathrm{U}}}\left( {\tilde {\mathbf{r}}} \right) & = {\mathbf{F}_{\mathrm{UB},k_\mathrm{U}}}{\left( {\tilde {\mathbf{r}}} \right)^H}{\mathbf{p}_{\mathrm{UB},k_\mathrm{U}}},\label{h_UB} \\ 
	{\mathbf{h}_{\mathrm{BD},k_\mathrm{D}}}\left( {\tilde {\mathbf{t}}} \right) & = {{\mathbf{G}_{\mathrm{BD},k_\mathrm{D}}}\left( {\tilde  {\mathbf{t}}} \right)}^H{\mathbf{p}_{\mathrm{BD},k_\mathrm{D}}},\label{h_BD} \\ 
	{\mathbf{h}_{\mathrm{BE},k_\mathrm{E}}}\left( {\tilde {\mathbf{t}}} \right) & = {{\mathbf{G}_{\mathrm{BE},k_\mathrm{E}}}\left( {\tilde  {\mathbf{t}}} \right)}^H{\mathbf{p}_{\mathrm{BE},k_\mathrm{E}}}, \label{h_BE}
\end{align}
where ${\mathbf{F}_{\mathrm{UB},k_\mathrm{U}}} \left( {\tilde {\mathbf{r}}} \right)\in {\mathbb{C}^{{L_{\mathrm{UB},k_\mathrm{U}}^\mathrm{r}} \times N_\mathrm{r}}}$ is the receive FRM for UL channel of UL user $k_\mathrm{U}$, ${\mathbf{G}_{\mathrm{BD},k_\mathrm{D}}} \left( {\tilde {\mathbf{t}}} \right)\in {\mathbb{C}^{{L_{\mathrm{BD},k_\mathrm{D}}^\mathrm{t}} \times N_\mathrm{t}}}$ and ${\mathbf{G}_{\mathrm{BE},k_\mathrm{E}}} \left( {\tilde {\mathbf{t}}} \right) \in {\mathbb{C}^{{L_{\mathrm{UE},k_\mathrm{E}}^\mathrm{t}} \times N_\mathrm{t}}}$ are the corresponding transmit FRMs for DL channels of DL user $k_\mathrm{D}$ and Eve $k_\mathrm{E}$. ${\mathbf{p}_{\mathrm{UB},k_\mathrm{U}}} \in {\mathbb{C}^{{L_{\mathrm{UB},k_\mathrm{U}}^\mathrm{r}} \times 1}}$ and ${\mathbf{p}_{\mathrm{BD},k_\mathrm{D}}} \in {\mathbb{C}^{{L_{\mathrm{BD},k_\mathrm{D}}^\mathrm{t}} \times 1}}$, ${\mathbf{p}_{\mathrm{BE},k_\mathrm{E}}} \in {\mathbb{C}^{{L_{\mathrm{BE},k_\mathrm{E}}^\mathrm{t}} \times 1}}$ are the path-response vectors (PRVs), which respectively represent the channel responses from UL user $k_\mathrm{U}$ to $O_\mathrm{r}$ at the BS and from $O_\mathrm{t}$ at the BS to DL user $k_\mathrm{D}$ and Eve $k_\mathrm{E}$. Since ${\mathbf{F}_{\mathrm{UB},k_\mathrm{U}}} \left( {\tilde {\mathbf{r}}} \right)$ and ${\mathbf{G}_{\mathrm{BD},k_\mathrm{D}}} \left( {\tilde {\mathbf{t}}} \right)$, ${\mathbf{G}_{\mathrm{BE},k_\mathrm{E}}} \left( {\tilde {\mathbf{t}}} \right)$ have the similar structures as ${\mathbf{F}_{\mathrm{SI}}} \left( {\tilde {\mathbf{r}}} \right)$ and ${\mathbf{G}_{\mathrm{SI}}} \left( {\tilde {\mathbf{t}}} \right)$, we can modify the calculations of ${\mathbf{F}_{\mathrm{SI}}} \left( {\tilde {\mathbf{r}}} \right)$ and ${\mathbf{G}_{\mathrm{SI}}} \left( {\tilde {\mathbf{t}}} \right)$ to obtain ${\mathbf{F}_{\mathrm{UB},k_\mathrm{U}}} \left( {\tilde {\mathbf{r}}} \right)$ and ${\mathbf{G}_{\mathrm{BD},k_\mathrm{D}}} \left( {\tilde {\mathbf{t}}} \right)$, ${\mathbf{G}_{\mathrm{BE},k_\mathrm{E}}} \left( {\tilde {\mathbf{t}}} \right)$ by replacing 
\begin{equation}
	\left\{ {L_\mathrm{SI}^\mathrm{r},\left\{ {\theta _{\mathrm{SI},i}^\mathrm{r}} \right\}_{i = 1}^{L_\mathrm{SI}^\mathrm{r}},\left\{ {\phi _{\mathrm{SI},i}^\mathrm{r}} \right\}_{i = 1}^{L_{\mathrm{SI}}^\mathrm{r}}} \right\} ,
\end{equation} 
with
\begin{equation}
	\left\{ {L_{\mathrm{UB},k_\mathrm{U}}^\mathrm{r},\left\{ {\theta _{\mathrm{UB},k_\mathrm{U},i}^\mathrm{r}} \right\}_{i = 1}^{L_{\mathrm{UB},k_\mathrm{U}}^\mathrm{r}},\left\{ {\phi _{\mathrm{UB},k_\mathrm{U},i}^\mathrm{r}} \right\}_{i = 1}^{L_{\mathrm{UB},k_\mathrm{U}}^\mathrm{r}}} \right\} ,
\end{equation} 
and replacing 
\begin{equation}
	\left\{ {L_\mathrm{SI}^\mathrm{t},\left\{ {\theta _{\mathrm{SI},i}^\mathrm{t}} \right\}_{i = 1}^{L_\mathrm{SI}^\mathrm{t}},\left\{ {\phi _{\mathrm{SI},i}^\mathrm{t}} \right\}_{i = 1}^{L_{\mathrm{SI}}^\mathrm{t}}} \right\} ,
\end{equation}
with 
\begin{align}
	& \left\{ {L_{\mathrm{BD},k_\mathrm{D}}^\mathrm{t},\left\{ {\theta _{\mathrm{BD},k_\mathrm{D},i}^\mathrm{t}} \right\}_{i = 1}^{L_{\mathrm{BD},k_\mathrm{D}}^\mathrm{t}},\left\{ {\phi _{\mathrm{BD},k_\mathrm{D},i}^\mathrm{t}} \right\}_{i = 1}^{L_{\mathrm{BD},k_\mathrm{D}}^\mathrm{t}}} \right\} , \\
	& \left\{ {L_{\mathrm{BE},k_\mathrm{E}}^\mathrm{t},\left\{ {\theta _{\mathrm{BE},k_\mathrm{E},i}^\mathrm{t}} \right\}_{i = 1}^{L_{\mathrm{BE},k_\mathrm{E}}^\mathrm{t}},\left\{ {\phi _{\mathrm{BE},k_\mathrm{E},i}^\mathrm{t}} \right\}_{i = 1}^{L_{\mathrm{BE},k_\mathrm{E}}^\mathrm{t}}} \right\}.
\end{align}
\subsection{Problem Formulation}
\begin{figure*}[!b]
	\textsc{\centering
		\hrulefill
		\begin{align}
			{\gamma _{k_\mathrm{U}}^{\mathrm{U}}} \nonumber
			& = \frac{{{{\left| {\mathbf{b}_{{k_\mathrm{U}}}^H {\mathbf{h}_{\mathrm{UB},{k_\mathrm{U}}}\left( \tilde {\mathbf{r}}\right)}} \right|}^2}{p_{{k_\mathrm{U}}}}}}{{\sum\limits_{i \in {\mathcal{K}_\mathrm{U}} \backslash \left\lbrace {k_\mathrm{U}}\right\rbrace } {{{\left| {\mathbf{b}_{{k_\mathrm{U}}}^H {\mathbf{h}_{\mathrm{UB},i}\left( \tilde {\mathbf{r}}\right)}} \right|}^2}} {p_i} + \rho \left( {{{\left| {\mathbf{b}_{{k_\mathrm{U}}}^H {\mathbf{H}_\mathrm{SI}\left( \tilde {\mathbf{t}}, \tilde {\mathbf{r}}\right)}\sum\limits_{{k_\mathrm{D}} \in {\mathcal{K}_\mathrm{D}}} {{\mathbf{w}_{{k_{\mathrm{D}}}}}} } \right|}^2} + {{\left| {\mathbf{b}_{{k_\mathrm{U}}}^H {\mathbf{H}_\mathrm{SI}\left( \tilde {\mathbf{t}}, \tilde {\mathbf{r}}\right)}\mathbf{v}} \right|}^2}} \right) + \left\| {{\mathbf{b}_{{k_\mathrm{U}}}}} \right\|_2^2\sigma _\mathrm{U}^2}} \nonumber\\
			& \mathop  = \limits^{\left( {{\alpha _1}} \right)} \frac{{{{\left| {{{\tilde h}_{\mathrm{UB},{k_\mathrm{U}},{k_\mathrm{U}}}}} \right|}^2}{p_{{k_\mathrm{U}}}}}}{{\sum\limits_{i \in {\mathcal{K}_\mathrm{U}} \backslash \left\lbrace {k_\mathrm{U}}\right\rbrace } {{{\left| {{{\tilde h}_{\mathrm{UB},{k_\mathrm{U}},i}}} \right|}^2}{p_i}}  + \sum\limits_{{k_\mathrm{D}} \in {\mathcal{K}_\mathrm{D}}} \mathrm{Tr}\left\{ { {{\mathbf{W}_{{k_\mathrm{D}}}}} {{\widetilde {\mathbf{H}}}_{\mathrm{SI},{k_\mathrm{U}}}}} \right\} + \mathrm{Tr}\left\{ {\mathbf{V}{{\widetilde {\mathbf{H}}}_{\mathrm{SI},{k_\mathrm{U}}}}} \right\}+ \left\| {{\mathbf{b}_{{k_\mathrm{U}}}}} \right\|_2^2\sigma _\mathrm{U}^2}} . \label{SINR_U} \\
			{\gamma _{k_\mathrm{D}}^{\mathrm{D}}} \nonumber
			& = \frac{{{{\left| {\mathbf{h}_{\mathrm{BD},{k_\mathrm{D}}}^H \left( \tilde {\mathbf{t}}\right) {\mathbf{w}_{{k_\mathrm{D}}}}} \right|}^2}}}{{\sum\limits_{i \in {\mathcal{K}_\mathrm{D}} \backslash \left\lbrace {k_\mathrm{D}}\right\rbrace } {{{\left| {\mathbf{h}_{\mathrm{BD},{k_\mathrm{D}}}^H \left( \tilde {\mathbf{t}}\right) {\mathbf{w}_i}} \right|}^2}}  + {{\left| {\mathbf{h}_{\mathrm{BD},{k_\mathrm{D}}}^H \left( \tilde {\mathbf{t}}\right) \mathbf{v}} \right|}^2} + \sum\limits_{{k_\mathrm{U}} \in {\mathcal{K}_\mathrm{U}}} {{{\left| {{h_{\mathrm{UD},{k_\mathrm{U}},{k_\mathrm{D}}}}} \right|}^2}{p_{{k_\mathrm{U}}}}}  + \sigma _{\mathrm{D},{k_\mathrm{D}}}^2}} \nonumber\\
			& \mathop  = \limits^{\left( {{\alpha _2}} \right)}   {\frac{{\mathrm{Tr}\left\{ {{\mathbf{W}_{{k_\mathrm{D}}}}{\mathbf{H}_{\mathrm{BD},{k_\mathrm{D}}}}} \right\}}}{{\sum\limits_{i \in {\mathcal{K}_\mathrm{D}} \backslash \left\lbrace {k_\mathrm{D}}\right\rbrace } {\mathrm{Tr}\left\{ {{\mathbf{W}_i}{\mathbf{H}_{\mathrm{BD},{k_\mathrm{D}}}}} \right\}}  + \mathrm{Tr}\left\{ {\mathbf{V}{\mathbf{H}_{\mathrm{BD},{k_\mathrm{D}}}}} \right\} + \sum\limits_{{k_\mathrm{U}} \in {\mathcal{K}_\mathrm{U}}} {{{\left| {{h_{\mathrm{UD},{k_\mathrm{U}},{k_\mathrm{D}}}}} \right|}^2}{p_{{k_\mathrm{U}}}}}  + \sigma _{\mathrm{D},{k_\mathrm{D}}}^2}}} . \label{SINR_D} \\
			\gamma _{k_\mathrm{U}}^{\mathrm{E}-\mathrm{U}} & = \sum\limits_{{k_\mathrm{E}} \in {\mathcal{K}_\mathrm{E}}} {\frac{{{{\left| {{h_{\mathrm{UE},{k_\mathrm{U}},{k_\mathrm{E}}}}} \right|}^2}{p_{{k_\mathrm{U}}}}}}{{{{\left| {\mathbf{h}_{\mathrm{BE},{k_\mathrm{E}}}^H\left( {\tilde {\mathbf{t}}} \right) \mathbf{v}} \right|}^2} + \sigma _{\mathrm{E},{k_\mathrm{E}}}^2}}} \mathop  = \limits^{\left( {{\alpha _3}} \right)} {\frac{{\sum\limits_{{k_\mathrm{E}} \in {\mathcal{K}_\mathrm{E}}} \left( {{{\left| {{h_{\mathrm{UE},{k_\mathrm{U}},{k_\mathrm{E}}}}} \right|}^2}{p_{{k_\mathrm{U}}}} \prod\limits_{i \in {\mathcal{K}_\mathrm{E}} \backslash \left\lbrace {k_\mathrm{E}}\right\rbrace } {\left( \mathrm{Tr}\left\{ {\mathbf{V}{\mathbf{H}_{\mathrm{BE},i}}} \right\} + \sigma _{\mathrm{E},i}^2\right)} }\right)  }}{{\prod\limits_{{k_\mathrm{E}} \in {\mathcal{K}_\mathrm{E}}} {\left( \mathrm{Tr}\left\{ {\mathbf{V}{\mathbf{H}_{\mathrm{BE},{k_\mathrm{E}}}}} \right\} + \sigma _{\mathrm{E},{k_\mathrm{E}}}^2\right)} }}} . \label{SINR_EU} \\
			\gamma _{k_\mathrm{D}}^{\mathrm{E}-\mathrm{D}} & = \sum\limits_{{k_\mathrm{E}} \in {\mathcal{K}_\mathrm{E}}} {\frac{{{{\left| {\mathbf{h}_{\mathrm{BE},{k_\mathrm{E}}}^H\left( {\tilde {\mathbf{t}}} \right){\mathbf{w}_{{k_\mathrm{D}}}}} \right|}^2}}}{{{{\left| {\mathbf{h}_{\mathrm{BE},{k_\mathrm{E}}}^H\left( {\tilde {\mathbf{t}}} \right) \mathbf{v}} \right|}^2} + \sigma _{\mathrm{E},{k_\mathrm{E}}}^2}}} \mathop  = \limits^{\left( {{\alpha _4}} \right)} {\frac{{\sum\limits_{{k_\mathrm{E}} \in {\mathcal{K}_\mathrm{E}}} \left( {\mathrm{Tr}\left\{ {{\mathbf{W}_{{k_\mathrm{D}}}}{\mathbf{H}_{\mathrm{BE},{k_\mathrm{E}}}}} \right\}\prod\limits_{i \in {\mathcal{K}_\mathrm{E}} \backslash \left\lbrace {k_\mathrm{E}}\right\rbrace } {\left( \mathrm{Tr}\left\{ {\mathbf{V}{\mathbf{H}_{\mathrm{BE},i}}} \right\} + \sigma _{\mathrm{E},i}^2 \right) } }\right)  }}{{\prod\limits_{{k_\mathrm{E}} \in {\mathcal{K}_\mathrm{E}}} {\left( \mathrm{Tr}\left\{ {\mathbf{V}{\mathbf{H}_{\mathrm{BE},{k_\mathrm{E}}}}} \right\} + \sigma _{\mathrm{E},{k_\mathrm{E}}}^2\right) } }}} . \label{SINR_ED}
	\end{align}}
\end{figure*}
The achievable rate of UL user $k_\mathrm{U}$ is given by $R_{{k_\mathrm{U}}}^\mathrm{U} = {\log _2}\left( {1 + \gamma _{{k_\mathrm{U}}}^\mathrm{U}} \right)$, where $\gamma _{{k_\mathrm{U}}}^\mathrm{U}$ is the receive signal-to-interference-plus-noise ratio (SINR) and given by \eqref{SINR_U} at the bottom of the page. Here, $\mathbf{b}_{{k_\mathrm{U}}} \in {\mathbb{C}^{{N_\mathrm{r}} \times 1}}$ is the receive beamformer of UL user $k_\mathrm{U}$, $0 < \rho  \ll 1$ is the SI loss coefficient representing the path loss and the SI cancellations in analog and digital domains. Besides, the achievable rate of DL user $k_\mathrm{D}$ is given by $R_{{k_\mathrm{D}}}^\mathrm{D} = {\log _2}\left( {1 + \gamma _{{k_\mathrm{D}}}^\mathrm{D}} \right)$, where $\gamma _{{k_\mathrm{D}}}^\mathrm{D}$ is the receive SINR and given by \eqref{SINR_D} at the bottom of the page.

To ensure secure communications, we consider a worst-case assumption. In particular, we assume that each Eve separately eavesdrops on UL and DL transmissions \cite{AN3} and can cancel all multi-user interference before decoding the information of a desired user \cite{AN4}. Furthermore, the $K_\mathrm{E}$ Eves aim to collaborate in processing the confidential information \cite{MA9}. Thus, under these assumptions, the achievable rates of Eves eavesdropping on UL user $k_\mathrm{U}$ and DL user $k_\mathrm{D}$ are respectively given by $R_{{k_\mathrm{U}}}^\mathrm{E-U} = {{\log }_2}\left( {1 + \gamma _{k_\mathrm{U}}^{\mathrm{E}-\mathrm{U}}} \right)$ and $R_{{k_\mathrm{D}}}^\mathrm{E-D} = {{\log }_2}\left( {1 + \gamma _{k_\mathrm{D}}^{\mathrm{E}-\mathrm{D}}} \right)$, where $\gamma _{k_\mathrm{U}}^{\mathrm{E}-\mathrm{U}}$ and $\gamma _{k_\mathrm{D}}^{\mathrm{E}-\mathrm{D}}$ are the receive SINRs and given by \eqref{SINR_EU} and \eqref{SINR_ED} at the bottom of the page. For the convenience of subsequent derivations, we rewrite the receive SINRs by the equalities marked by $\left( {\alpha _1}\right)$-$\left( {\alpha _4}\right) $ in \eqref{SINR_U}-\eqref{SINR_ED}, where 
\begin{align}
	&\mathbf{W}_{k_\mathrm{D}}   = \mathbf{w}_{k_\mathrm{D}}\mathbf{w}_{k_\mathrm{D}}^H \in {\mathbb{C}^{{N_\mathrm{t}} \times {N_\mathrm{t}}}}, \\
	&\mathbf{V}  = \mathbf{vv}^H \in {\mathbb{C}^{{N_\mathrm{t}} \times {N_\mathrm{t}}}}, \\
	&{{\tilde h}_{\mathrm{UB},{k_\mathrm{U}},i}}  = \mathbf{b}_{{k_\mathrm{U}}}^H {\mathbf{h}_{\mathrm{UB},i}\left( \tilde {\mathbf{r}}\right)} \in {\mathbb{C}^{1 \times 1}}, \\
	&{{\tilde {\mathbf{h}}}_{\mathrm{SI},{k_\mathrm{U}}}}  = \sqrt \rho {\mathbf{H}_\mathrm{SI}^H} \left( {\tilde {\mathbf{t}},\tilde {\mathbf{r}}} \right)\mathbf{b}_{{k_\mathrm{U}}} \in {\mathbb{C}^{{N_\mathrm{t}} \times 1}}, \\ &{\widetilde {\mathbf{H}}_{\mathrm{SI},{k_\mathrm{U}}}}  = {{\tilde {\mathbf{h}}}_{\mathrm{SI},{k_\mathrm{U}}}}\tilde {\mathbf{h}}_{\mathrm{SI},{k_\mathrm{U}}}^H \in {\mathbb{C}^{{N_\mathrm{t}} \times {N_\mathrm{t}}}}, \\
	&{\mathbf{H}_{\mathrm{BD},k_\mathrm{D}}}  = {\mathbf{h}_{\mathrm{BD},k_\mathrm{D}}\left( {\tilde {\mathbf{t}}} \right)}\mathbf{h}_{\mathrm{BD},k_\mathrm{D}}^H\left( {\tilde {\mathbf{t}}} \right)\in {\mathbb{C}^{{N_\mathrm{t}} \times {N_\mathrm{t}}}}, \\
	&{\mathbf{H}_{\mathrm{BE},k_\mathrm{E}}}  = {\mathbf{h}_{\mathrm{BE},k_\mathrm{E}}\left( {\tilde {\mathbf{t}}} \right)}\mathbf{h}_{\mathrm{BE},k_\mathrm{E}}^H\left( {\tilde {\mathbf{t}}} \right)\in {\mathbb{C}^{{N_\mathrm{t}} \times {N_\mathrm{t}}}}.
\end{align}

In the paper, we focus on maximizing the SSR\footnote{The goal of this paper is to provide a performance upper bound for realistic scenarios and robust designs. Thus, we assume the perfect CSI of all channels is available at the BS for channel response and SSR calculations. Despite the challenges in acquiring perfect CSI, existing works \cite{MA14,MA15,AN4} have proposed some practical methods that can achieve satisfactory CSI estimation for MA-aided or FD-based systems. Besides, the impact of imperfect CSI on the considered system will be evaluated via simulations in Section \ref{per_section}.} of the users, i.e.,
\begin{equation}
	R_\mathrm{SSR} =  \sum\limits_{{k_\mathrm{U}} \in {\mathcal{K}_\mathrm{U}}} {\left[ R_{{k_\mathrm{U}}}^\mathrm{U} - R_{{k_\mathrm{U}}}^\mathrm{E-U} \right] ^+}  + \sum\limits_{{k_\mathrm{D}} \in {\mathcal{K}_\mathrm{D}}} {\left[ R_{{k_\mathrm{D}}}^\mathrm{D} - R_{{k_\mathrm{D}}}^\mathrm{E-D} \right] ^+} ,
\end{equation}
by jointly optimizing the MA positions, $\tilde {\mathbf{t}}$ and $\tilde {\mathbf{r}}$, the transmit, AN, and receive beamformers at the BS, $ {\mathbf{w}_{{k_\mathrm{D}}}}$, $\mathbf{v}$, and $\mathbf{b}_{k_\mathrm{U}}$, and the UL powers, ${p_{{k_\mathrm{U}}}}$. The corresponding optimization problem is formulated as follows\footnote{This paper designs the resource allocation algorithm based on an SSR maximization problem. We note that the derived solutions can also be applied to minimize the total UL and DL transmit powers with appropriate rate constraints due to the duality between rate and power optimization \cite{AN4}.}.
\begin{align} \label{max1}
	& \mathop {\mathrm{maximize} }\limits_{\tilde{\mathbf{t}},\tilde{\mathbf{r}},{\mathbf{w}_{{k_\mathrm{D}}}}, \mathbf{v},{p_{{k_\mathrm{U}}}}, {\mathbf{b}_{{k_\mathrm{U}}}} } \quad {R_\mathrm{SSR}} \\
	&\mathrm{s.t.} \quad \text{C1}: \left\| {{\mathbf{b}_{{k_\mathrm{U}}}}} \right\|_2^2 = 1 , \ \forall {k_\mathrm{U}} \in \mathcal{K}_\mathrm{U}, \nonumber\\
	&\hspace{2.3em} \text{C2}: \sum\limits_{{k_\mathrm{D}} \in {\mathcal{K}_\mathrm{D}}} {\mathrm{Tr}\left\{ \mathbf{w}_{k_\mathrm{D}}\mathbf{w}_{k_\mathrm{D}}^H \right\}}  + \mathrm{Tr}\left\{ \mathbf{v} \mathbf{v}^H \right\} \le P_{\max }^\mathrm{D}, \nonumber\\
	&\hspace{2.3em} \text{C3}: 0 \le {p_{{k_\mathrm{U}}}} \le P_{\max ,{k_\mathrm{U}}}^\mathrm{U} , \ \forall {k_\mathrm{U}} \in \mathcal{K}_\mathrm{U}, \nonumber\\
	&\hspace{2.3em} \text{C4}: \tilde {\mathbf{t}} \in {\mathcal{C}_\mathrm{t}}, \ \tilde {\mathbf{r}} \in {\mathcal{C}_\mathrm{r}}, \nonumber\\
	&\hspace{2.3em} \text{C5}: {\left\| {{\mathbf{t}_a} - {\mathbf{t}_{\tilde a}}} \right\|_2} \ge D , \ 1 \le a \ne {\tilde a} \le {N_\mathrm{t}}, \nonumber\\
	&\hspace{2.3em} \text{C6}: {\left\| {{\mathbf{r}_b} - {\mathbf{r}_{\tilde b}}} \right\|_2} \ge D,\ 1 \le b \ne {\tilde b} \le {N_\mathrm{r}}. \nonumber
\end{align}
Here, constraint C1 normalizes the receive beamformer. $P_{\max }^\mathrm{D} > 0$ and $P_{\max ,{k_\mathrm{U}}}^\mathrm{U} > 0$ in constraints C2 and C3 are the maximum transmit powers of the BS and UL user ${k_\mathrm{U}}$, respectively. Constraint C4 limits the ranges of MA movements. Constraints C5 and C6 ensure that minimum inter-MA distance $D$ at the BS for practical implementation. The operator ${\left[  \cdot  \right]^ + }$ has no impact on the optimization and will be omitted in the subsequent derivations.

Note that problem\;\eqref{max1} is a highly non-convex optimization problem. Specifically, the non-convexity of the objective function and the minimum inter-MA distance constraints C5 and C6, along with the couplings between the optimization variables, make the optimization problem particularly intractable. To the best of our knowledge, existing optimization tools cannot be directly applied to obtain the globally optimal solution. Thus, we propose an AO algorithm to solve problem\;\eqref{max1} in the next section.
\section{Proposed Solution} \label{3}
In this section, we propose an AO algorithm to address problem \eqref{max1}. Indeed, AO is a widely applicable methodology that decomposes the original problem into several sub-problems and iteratively solves each one while holding the optimization variables in other sub-problems fixed \cite{MA3,MA11,AN2,AN4}. Specifically, we decompose problem\;\eqref{max1} into three sub-problems, i.e., iteratively optimizing $\left\{ {\tilde {\mathbf{t}},\tilde {\mathbf{r}}} \right\}$, $\left\{ {{\mathbf{w}_{{k_\mathrm{D}}}},\mathbf{v},{p_{{k_\mathrm{U}}}}} \right\}$, and ${\mathbf{b}_{{k_\mathrm{U}}}} $.
\subsection{Sub-Problem 1: Optimize $\left\{ {\tilde {\mathbf{t}},\tilde {\mathbf{r}}} \right\}$ With Given $\left\{ {{\mathbf{w}_{{k_\mathrm{D}}}},\mathbf{v},{p_{{k_\mathrm{U}}}}} \right\}$ and $ {\mathbf{b}_{{k_\mathrm{U}}}} $}
With the given $\left\{ {{\mathbf{w}_{{k_\mathrm{D}}}},\mathbf{v},{p_{{k_\mathrm{U}}}}} \right\}$ and $ {\mathbf{b}_{{k_\mathrm{U}}}} $, the SSR can be expressed as a function of $\tilde {\mathbf{t}}$ and $\tilde {\mathbf{r}}$.  Therefore, sub-problem\;1 can be formulated as
\begin{align} \label{max2}
	& \mathop {\mathrm{maximize} }\limits_{\tilde{\mathbf{t}},\tilde{\mathbf{r}}} \quad
	{R_\mathrm{SSR} \left( \tilde{\mathbf{t}}, \tilde{\mathbf{r}} \right)}\\
	& \mathrm{s.t.} \quad {\text{C4, C5, C6}}. \nonumber
\end{align}

The conventional alternating position optimization (APO) \cite{MA4}, which iteratively fixes the other MAs while moving only one, may converge to an undesired local optimal solution because the given positions of other MAs narrow the optimization space of the current MA to a tiny region \cite{MA4}. To address this problem, we propose the MVPSO algorithm, which is an effective improvement of the standard PSO \cite{Ding1,Ding2} by replacing the single velocity of each particle in the iterations with multiple candidate velocities, to simultaneously optimize the positions of all transmit and receive MAs. The details of MVPSO are presented below.
\begin{algorithm}[!t]
	\caption{Multi-Velocity Particle Swarm Optimization}
	\label{MVPSO}
	\small
	\renewcommand{\algorithmicrequire}{\textbf{Initialization:}}
	\renewcommand{\algorithmicensure}{\textbf{Output:}}
	\begin{algorithmic}[1]
		\REQUIRE Set initial MVPSO parameters, i.e., $N$, $Q$, $A$, $D$, $\tau_\mathrm{t}$, $\tau_\mathrm{r}$, $\bar I$, $\bar J$, $c_1$, and $c_2$.
		\ENSURE $\mathbf{u}$.
		\STATE Initialize the positions and velocities of $N$ particles as $\mathbf{U}^{(0)}$ and $\mathbf{Z}^{(0)}$, respectively;
		\STATE Calculate the fitness value of each particle by \eqref{fitness};
		\STATE Initialize the personal best position $\mathbf{u}_{\mathrm{pbest},n} = \mathbf{u}_n^{(0)}$ and the global best position $\mathbf{u}_\mathrm{gbest} = \arg \mathop {\max }\limits_{\mathbf{u}_n^{(0)}} \left\{  {\mathcal{F}\left( {\mathbf{u}_1^{(0)}} \right), \cdots, \mathcal{F}\left( {\mathbf{u}_N^{(0)}} \right)} \right\}$;
		\FOR{$q=1:1:Q$}
		\FOR{$n=1:1:N$}
		\STATE Initialize the velocity components matrix ${\mathbf{\Psi}_n ^{\left( q \right)}}$;
		\FOR{$\bar i=1:1:\bar I$}
		\STATE Initialize the combination weight vector ${\mathbf{c}_{n,\bar i}}$;
		\STATE Calculate the candidate velocity $\mathbf{z}_{n,\bar i}^{\left( q \right)}$ by \eqref{multi-v} and the candidate position $\mathbf{u}_{n,\bar i}^{\left( q \right)}$ by \eqref{position};
		\STATE Calculate the fitness value of each candidate position by \eqref{fitness};
		\ENDFOR
		\STATE Update the position ${\mathbf{u}_n^{(q)}}$ by \eqref{pos} and the corresponding velocity $\mathbf{z}_n^{\left( q \right)}$ by \eqref{vel};
		\IF{$\mathcal{F}\left( {\mathbf{u}_n^{(q)}} \right) > \mathcal{F}\left({\mathbf{u}_{\mathrm{pbest},n}}\right)$}
		\STATE Update $\mathbf{u}_{\mathrm{pbest},n} = \mathbf{u}_n^{(q)}$;
		\ENDIF
		\IF{$\mathcal{F}\left( {\mathbf{u}_n^{(q)}} \right) > \mathcal{F}\left({\mathbf{u}_\mathrm{gbest}}\right)$}
		\STATE Update $\mathbf{u}_\mathrm{gbest} = {\mathbf{u}_n^{(q)}}$;
		\ENDIF
		\ENDFOR
		\ENDFOR
		\RETURN $\mathbf{u} = \mathbf{u}_\mathrm{gbest}$.
	\end{algorithmic}
\end{algorithm}

We first randomly initialize the positions and velocities of $N$ particles as ${\mathbf{U}^{\left( 0 \right)}} = \left\{ {\mathbf{u}_1^{\left( 0 \right)}, \ldots ,\mathbf{u}_N^{\left( 0 \right)}} \right\} \in {\mathbb{R}^{2\left( N_\mathrm{t}+N_\mathrm{r}\right)  \times N}}$ and ${\mathbf{Z}^{\left( 0 \right)}} = \left\{ {\mathbf{z}_1^{\left( 0 \right)}, \ldots ,\mathbf{z}_N^{\left( 0 \right)}} \right\} \in {\mathbb{R}^{2\left( N_\mathrm{t}+N_\mathrm{r}\right)  \times N}}$, respectively, where each particle's position represents a possible solution for the antenna position vector, i.e., $\mathbf{u}_n^{\left( 0 \right)} = {\left[ {{{\tilde {\mathbf{t}}}_n^{\left( 0 \right)T}},{{\tilde {\mathbf{r}}}_n^{\left( 0 \right)T}}} \right]^T} \in {\mathbb{R}^{2\left( N_\mathrm{t}+N_\mathrm{r}\right)  \times 1}}$ ($1 \le n \le N$). Without loss of generality, we assume that each moving region is a square with size $A \times A$. Each element in ${\mathbf{U}^{\left( 0 \right)}}$ obeys the uniform distribution over the real-number interval $\left[ { - \frac{A}{2},\frac{A}{2}} \right]$ to ensure that the initial positions of MAs do not exceed the corresponding moving regions, i.e., constraint C4 holds. Then, the personal best position of the $n$-th particle $\mathbf{u}_{\mathrm{pbest},n}$ are initialized as $\mathbf{u}_n^{\left( 0 \right)}$ and the global best position $\mathbf{u}_{\mathrm{gbest}}$ is selected based on the fitness function. After completing the initialization, the processing procedures of the MVPSO algorithm are summarized in Algorithm\;\ref{MVPSO}. Let $Q$ denote the maximum number of iterations, the introduction of Algorithm\;\ref{MVPSO} is given as follows.

\subsubsection{Define Fitness Function}
Considering constraints C5 and C6 on the minimum inter-MA distance, we first define a penalty function as
\begin{align}\label{penalty_function}
	\mathcal{P}\left( {\mathbf{u}_n^{\left( q \right)}} \right) 
	& = \tau_\mathrm{t} \sum\limits_{a = 1}^{{N_\mathrm{t}} - 1} {\sum\limits_{\tilde a = a + 1}^{{N_\mathrm{t}}} {\delta \left( {{{\left\| {{\mathbf{t}^{\left( q \right)}_a} - {\mathbf{t}^{\left( q \right)}_{\tilde a}}} \right\|}_2} < D} \right)} } \nonumber\\
	& + \tau_\mathrm{r} \sum\limits_{b = 1}^{{N_\mathrm{r}} - 1} {\sum\limits_{\tilde b = b + 1}^{{N_\mathrm{r}}} {\delta \left( {{{\left\| {{\mathbf{r}^{\left( q \right)}_b} - {\mathbf{r}^{\left( q \right)}_{\tilde b}}} \right\|}_2} < D} \right)}} ,
\end{align}
where ${\mathbf{u}_n^{\left( q \right)}}$ is the position of the $n$-th particle in the $q$-th ($1 \le q \le Q$) iteration. ${\delta \left(  \cdot  \right)}$ is an indicator function, equaling 1 when the condition within the bracket is true; otherwise, it equals 0. $\tau_\mathrm{t}$ and $\tau_\mathrm{r}$ are the positive penalty factors utilized to regulate the severity of the penalty. Assume that the best position has the largest fitness value. Based on this given penalty function, for maximizing the SSR, the fitness function is defined as
\begin{equation} \label{fitness}
	\mathcal{F}\left( {\mathbf{u}_n^{\left( q \right)}} \right) = {R_{\mathrm{SSR} }}\left( {\mathbf{u}_n^{\left( q \right)}} \right) - \mathcal{P}\left( {\mathbf{u}_n^{\left( q \right)}} \right),
\end{equation}
where the values of $\tau_\mathrm{t}$ and $\tau_\mathrm{r}$ consistently maintain ${R_{\mathrm{SSR} }}\left( {\mathbf{u}_n^{\left( q \right)}} \right) - \left( \tau_\mathrm{t}+\tau_\mathrm{r}\right) < 0$. Thus, the penalty function can push the particles to satisfy the minimum inter-MA distance. In other words, with the progression of iterations, $\mathcal{P}\left( {\mathbf{u}_n^{\left( q \right)}} \right)$ will converge to zero.
\subsubsection{Update Positions and Velocities}\label{multi_v}
The candidate positions of the $n$-th particle in the $q$-th iteration are updated by ${\bar I}$ candidate velocities, i.e., 
\begin{equation}\label{position}
	\mathbf{u}_{n,\bar i}^{\left( q \right)} = \mathcal{B}\left\{ {\mathbf{u}_n^{\left( {q - 1} \right)} + \mathbf{z}_{n,\bar i}^{\left( q \right)}} \right\} ,
\end{equation}
where $\mathbf{z}_{n,\bar i}^{\left( q \right)}$ is the $\bar i$-th ($0< \bar i < \bar I$) candidate velocity, which will be specified later. $\mathcal{B} \left\lbrace \mathbf{u}\right\rbrace $ is a function, that projects each entry of vector $\mathbf{u}$ to the nearest boundary if it exceeds the feasible region, to satisfy constraint C4, i.e., 
\begin{equation}
{\left[ {\mathcal{B}\left\{ \mathbf{u} \right\}} \right]_i} = \left\{
\begin{array}{rcl}
	\frac{A}{2}, & \mathrm{if} & {{\left[ \mathbf{u} \right]_i}>\frac{A}{2}} ,\\
	{\left[ \mathbf{u} \right]_i}, & \mathrm{if} & {-\frac{A}{2} \leq {\left[ \mathbf{u} \right]_i} \leq \frac{A}{2}} , \\
	-\frac{A}{2}, & \mathrm{if} & {{\left[ \mathbf{u} \right]_i} < -\frac{A}{2}} .\\
\end{array} \right.
\end{equation}

For standard PSO, the inertia weight in each particle's velocity decreases with the number of iterations in the interval $\left[\omega_\mathrm{min},\omega_\mathrm{max}\right] $, i.e.,
\begin{equation}\label{w}
	\omega  = {\omega _{\max }} - \frac{{\left( {{\omega _{\max }} - {\omega _{\min }}} \right)q}}{Q} .
\end{equation}
Generally, a small $\omega$ leads to local exploitation for optimal solutions within the current region, whereas a large $\omega$ signifies that the particles can globally explore to evade undesired local optimal solutions \cite{PSO}. Hence, the standard PSO lacks exploitation in the early iterations and exploration in the late iterations. Based on the aforementioned observations, we modify the standard PSO and propose the MVPSO, in which each particle can select the optimal velocity from multiple candidate velocities in each iteration. 

The ${\bar I}$ candidate velocities in \eqref{position} are generated by the weighted combinations of ${\bar J}$ velocity components, i.e., 
\begin{equation}\label{multi-v}
	\mathbf{z}_{n,\bar i}^{\left( q \right)} = {\mathbf{\Psi}_n ^{\left( q \right)}}{\mathbf{c}_{n,\bar i}} ,
\end{equation}
where ${\mathbf{c}_{n,\bar i}} \in {\mathbb{R}^{\bar J  \times 1}}$ is a constant vector of the $n$-th particle, which represents the combination weights for generating the $\bar i$-th candidate velocity. ${\mathbf{\Psi}_n ^{\left( q \right)}} = \left[ {\boldsymbol{\psi} _{n,1}^{\left( q \right)}, \ldots ,\boldsymbol{\psi}_{n,\bar J}^{\left( q \right)}} \right] \in {\mathbb{R}^{2\left( N_\mathrm{t}+N_\mathrm{r}\right)  \times \bar J}}$ is the collection of velocity components for the $n$-th particle and $\boldsymbol{\psi}_{n,\bar j}^{\left( q \right)} \in {\mathbb{R}^{2\left( N_\mathrm{t}+N_\mathrm{r}\right)  \times 1}}$ represents the $\bar j$-th ($1 \le  \bar j \le  \bar J$) velocity component, which is calculated as
\begin{align}
	\boldsymbol{\psi}_{n,\bar j}^{\left( q \right)} = \omega_{\bar j} \mathbf{z}_n^{\left( {q - 1} \right)} \nonumber
	& + {c_1}{\mathbf{e}_1} \odot \left( {{\mathbf{u}_{\mathrm{pbest},n}} - \mathbf{u}_n^{\left( {q - 1} \right)}} \right) \nonumber\\
	& + {c_2}{\mathbf{e}_2} \odot \left( {{\mathbf{u}_\mathrm{gbest}} - \mathbf{u}_n^{\left( {q - 1} \right)}} \right) ,
\end{align}
where $\mathbf{z}_n^{\left( {q - 1} \right)}$ is the selected optimal velocity in the $\left( q-1\right) $-th iteration which will be specified later. $\omega_{\bar j}$ is the inertia weight of the $\bar j$-th velocity component. $c_1$ and $c_2$ are the personal and global learning factors that push each particle toward the personal and global best positions, respectively. To reduce the possibility of converging to an undesired local optimal solution, two random vectors $\mathbf{e}_1$ and $\mathbf{e}_2$, with uniformly distributed entries in the range $\left[0,1\right]$, are utilized. 

In general, the combination weights of different velocity components can be configured to simultaneously accommodate exploitation and exploration. However, the velocity components in ${\mathbf{\Psi}_n ^{\left( q \right)}}$ may not cover the optimal velocity but contain the velocity components with biased local and biased global search behaviors. In other words, both the $\bar J$ velocity components themselves and their weighted combinations can serve as the candidate velocities. Thus, the constant vector ${\mathbf{c}_{n,\bar i}}$ is introduced to control the combination of the velocity components. It is worth noting that the standard PSO can be regarded as a special case of the proposed MVPSO by only setting one candidate velocity. Similar multi-candidate approaches have been employed to address optimization problems in wireless communications, e.g., \cite{MA5}.

Finally, the position of the $n$-th particle in the $q$-th iteration is selected from $\bar I$ candidate solutions which can achieve the maximum fitness value in \eqref{fitness}, i.e.,
\begin{equation}\label{pos}
	\mathbf{u}_n^{\left( q \right)} = \mathop {\arg \max }\limits_{\mathbf{u}_{n,\bar i}^{\left( q \right)}} \left\{ {\mathcal{F}\left( {\mathbf{u}_{n,1}^{\left( q \right)}} \right), \ldots ,\mathcal{F}\left( {\mathbf{u}_{n,\bar I}^{\left( q \right)}} \right)} \right\} .
\end{equation}
The corresponding velocity is updated to the candidate velocity associated with the selected solution, i.e.,
\begin{equation}\label{vel}
	\mathbf{z}_n^{\left( q \right)} = \left\{ {\mathbf{z}_{n,\bar i}^{\left( q \right)}\left| {\mathcal{B}\left\{ {\mathbf{u}_n^{\left( {q - 1} \right)} + \mathbf{z}_{n,\bar i}^{\left( q \right)}} \right\} = \mathbf{u}_n^{\left( q \right)}} \right.} \right\} ,
\end{equation}
The generation and selection of each particle's position and velocity are presented in lines 6-12.
\subsubsection{Update Personal and Global Best Positions}
After obtaining the particles' positions, the personal and global best positions are updated if the fitness value at the current position exceeds the personal and global best fitness values, respectively. The corresponding pseudo-code is shown in lines 13-18. After $Q$ iterations, an optimized solution for the antenna position vector is obtained by the global best position, i.e., line\;21.
\subsection{Sub-Problem 2: Optimize $\left\{ {{\mathbf{w}_{{k_\mathrm{D}}}},\mathbf{v},{p_{{k_\mathrm{U}}}}} \right\}$ With Given $\left\{ {\tilde {\mathbf{t}},\tilde {\mathbf{r}}} \right\}$ and $ {\mathbf{b}_{{k_\mathrm{U}}}} $}\label{opt_wv}
Based on the rule of the logarithmic function, we define $f$ and $g$, which are given by \eqref{f} and \eqref{g} at the bottom of the page, and rewrite the SSR as $R_\mathrm{SSR} = f-g $.
\begin{figure*}[!b]
	\textsc{\centering
		\hrulefill
		\begin{align}
			f & = \sum\limits_{{k_\mathrm{D}} \in {\mathcal{K}_\mathrm{D}}} {{{\log }_2}\left( {{{\sum\limits_{i\in {\mathcal{K}_\mathrm{D}}} {\mathrm{Tr}\left\{ {{\mathbf{W}_i}{\mathbf{H}_{\mathrm{BD},{k_\mathrm{D}}}}} \right\}}  + \mathrm{Tr}\left\{ {\mathbf{V}{\mathbf{H}_{\mathrm{BD},{k_\mathrm{D}}}}} \right\} + \sum\limits_{{k_\mathrm{U}} \in {\mathcal{K}_\mathrm{U}}} {{{\left| {{h_{\mathrm{UD},{k_\mathrm{U}},{k_\mathrm{D}}}}} \right|}^2}{p_{{k_\mathrm{U}}}}}  + \sigma _{\mathrm{D},{k_\mathrm{D}}}^2}}} \right)} \nonumber\\
			& + \sum\limits_{{k_\mathrm{U}} \in {\mathcal{K}_\mathrm{U}}} {{{\log }_2}\left( {{\sum\limits_{i \in {\mathcal{K}_\mathrm{U}} } {{{\left| {{{\tilde h}_{\mathrm{UB},{k_\mathrm{U}},i}}} \right|}^2}{p_i}}  + \sum\limits_{{k_\mathrm{D}} \in {\mathcal{K}_\mathrm{D}}} \mathrm{Tr}\left\{ { {{\mathbf{W}_{{k_\mathrm{D}}}}} {{\widetilde {\mathbf{H}}}_{\mathrm{SI},{k_\mathrm{U}}}}} \right\} + \mathrm{Tr}\left\{ {\mathbf{V}{{\widetilde {\mathbf{H}}}_{\mathrm{SI},{k_\mathrm{U}}}}} \right\}+ \left\| {{\mathbf{b}_{{k_\mathrm{U}}}\left(\tilde {\mathbf{r}}\right)}} \right\|_2^2\sigma _\mathrm{U}^2}} \right)} \nonumber\\
			& + \left( K_\mathrm{D}+K_\mathrm{U} \right)  {{{\log }_2}\left( {{\prod\limits_{{k_\mathrm{E}} \in {\mathcal{K}_\mathrm{E}}} {\left( \mathrm{Tr}\left\{ {\mathbf{V}{\mathbf{H}_{\mathrm{BE},{k_\mathrm{E}}}}} \right\} + \sigma _{\mathrm{E},{k_\mathrm{E}}}^2\right) } }} \right)} . \label{f} \\
			g & = \sum\limits_{{k_\mathrm{D}} \in {\mathcal{K}_\mathrm{D}}} {{{\log }_2}\left( {{{\sum\limits_{i \in {\mathcal{K}_\mathrm{D}} \backslash \left\lbrace {k_\mathrm{D}}\right\rbrace } {\mathrm{Tr}\left\{ {{\mathbf{W}_i}{\mathbf{H}_{\mathrm{BD},{k_\mathrm{D}}}}} \right\}}  + \mathrm{Tr}\left\{ {\mathbf{V}{\mathbf{H}_{\mathrm{BD},{k_\mathrm{D}}}}} \right\} + \sum\limits_{{k_\mathrm{U}} \in {\mathcal{K}_\mathrm{U}}} {{{\left| {{h_{\mathrm{UD},{k_\mathrm{U}},{k_\mathrm{D}}}}} \right|}^2}{p_{{k_\mathrm{U}}}}}  + \sigma _{\mathrm{D},{k_\mathrm{D}}}^2}}} \right)} \nonumber\\
			& + \sum\limits_{{k_\mathrm{D}} \in {\mathcal{K}_\mathrm{D}}} {{{\log }_2}\left( {{\prod\limits_{{k_\mathrm{E}} \in {\mathcal{K}_\mathrm{E}}} {\left( \mathrm{Tr}\left\{ {\mathbf{V}{\mathbf{H}_{\mathrm{BE},{k_\mathrm{E}}}}} \right\} + \sigma _{\mathrm{E},{k_\mathrm{E}}}^2\right) } }} + {{\sum\limits_{{k_\mathrm{E}} \in {\mathcal{K}_\mathrm{E}}} \left( {\mathrm{Tr}\left\{ {{\mathbf{W}_{{k_\mathrm{D}}}}{\mathbf{H}_{\mathrm{BE},{k_\mathrm{E}}}}} \right\}\prod\limits_{i \in {\mathcal{K}_\mathrm{E}}\backslash \left\lbrace {k_\mathrm{E}}\right\rbrace } {\left( \mathrm{Tr}\left\{ {\mathbf{V}{\mathbf{H}_{\mathrm{BE},i}}} \right\} + \sigma _{\mathrm{E},i}^2 \right) } }\right)  }} \right)} \nonumber\\
			& + \sum\limits_{{k_\mathrm{U}} \in {\mathcal{K}_\mathrm{U}}} {{{\log }_2}\left( {{\sum\limits_{i \in {\mathcal{K}_\mathrm{U}}\backslash\left\lbrace {k_\mathrm{U}}\right\rbrace } {{{\left| {{{\tilde h}_{\mathrm{UB},{k_\mathrm{U}},i}}} \right|}^2}{p_i}}  + \sum\limits_{{k_\mathrm{D}} \in {\mathcal{K}_\mathrm{D}}} \mathrm{Tr}\left\{ { {{\mathbf{W}_{{k_\mathrm{D}}}}} {{\widetilde {\mathbf{H}}}_{\mathrm{SI},{k_\mathrm{U}}}}} \right\} + \mathrm{Tr}\left\{ {\mathbf{V}{{\widetilde {\mathbf{H}}}_{\mathrm{SI},{k_\mathrm{U}}}}} \right\}+ \left\| {{\mathbf{b}_{{k_\mathrm{U}}}\left(\tilde {\mathbf{r}}\right)}} \right\|_2^2\sigma _\mathrm{U}^2}} \right)} \nonumber\\
			& + \sum\limits_{{k_\mathrm{U}} \in {\mathcal{K}_\mathrm{U}}} {{{\log }_2}\left( {{\prod\limits_{{k_\mathrm{E}} \in {\mathcal{K}_\mathrm{E}}} {\left( \mathrm{Tr}\left\{ {\mathbf{V}{\mathbf{H}_{\mathrm{BE},{k_\mathrm{E}}}}} \right\} + \sigma _{\mathrm{E},{k_\mathrm{E}}}^2\right)} }} + {{\sum\limits_{{k_\mathrm{E}} \in {\mathcal{K}_\mathrm{E}}} \left( {{{\left| {{h_{\mathrm{UE},{k_\mathrm{U}},{k_\mathrm{E}}}}} \right|}^2}{p_{{k_\mathrm{U}}}} \prod\limits_{i \in {\mathcal{K}_\mathrm{E}}\backslash\left\lbrace {k_\mathrm{E}}\right\rbrace } {\left( \mathrm{Tr}\left\{ {\mathbf{V}{\mathbf{H}_{\mathrm{BE},i}}} \right\} + \sigma _{\mathrm{E},i}^2\right)} }\right)  }} \right)} .\label{g}
		\end{align}}
\end{figure*}
Thus, with given $\left\{ {\tilde {\mathbf{t}},\tilde {\mathbf{r}}} \right\}$ and $ {\mathbf{b}_{{k_\mathrm{U}}}}$, sub-problem\;2 can be formulated as
\begin{align}\label{max3}
	& \mathop {\mathrm{maximize} }\limits_{{\mathbf{W}_{{k_\mathrm{D}}}}, \mathbf{V},{p_{{k_\mathrm{U}}}} } \quad f-g \\
	&\mathrm{s.t.} \quad \text{C3, C7}: \sum\limits_{{k_\mathrm{D}} \in {\mathcal{K}_\mathrm{D}}} {\mathrm{Tr}\left\{ \mathbf{W}_{k_\mathrm{D}} \right\}}  + \mathrm{Tr}\left\{ \mathbf{V}\right\} \le P_{\max }^\mathrm{D},\nonumber\\
	&\hspace{2.3em} \text{C8}: {\mathbf{W}_{k_\mathrm{D}}} \succeq \mathbf{0}, \mathbf{V} \succeq \mathbf{0}, \ \forall {k_\mathrm{D}} \in \mathcal{K}_\mathrm{D}, \nonumber\\
	&\hspace{2.3em} \text{C9}: \mathrm{Rank}\left\lbrace {\mathbf{W}_{k_\mathrm{D}}} \right\rbrace  \le 1, \mathrm{Rank}\left\lbrace  \mathbf{V} \right\rbrace  \le 1, \ \forall {k_\mathrm{D}} \in \mathcal{K}_\mathrm{D}.  \nonumber
\end{align}
\begin{figure*}[!b]
	\textsc{\centering
		\hrulefill
		\begin{align}\label{g_est}
			& {g} \left( {{\mathbf{W}_{{k_\mathrm{D}}}},\mathbf{V},{p_{{k_\mathrm{U}}}}} \right)  \le {g}\left( {\mathbf{W}_{{k_\mathrm{D}}}^{\left( m \right)},{\mathbf{V}^{\left( m \right)}},p_{{k_\mathrm{U}}}^{\left( m \right)}} \right) + \sum\limits_{{k_\mathrm{D}} \in {\mathcal{K}_\mathrm{D}}} {\mathrm{Tr}\left\{ {{{\left( {{\nabla _{{\mathbf{W}_{{k_\mathrm{D}}}}}}{g}\left( {\mathbf{W}_{{k_\mathrm{D}}}^{\left( m \right)},{\mathbf{V}^{\left( m \right)}},p_{{k_\mathrm{U}}}^{\left( m \right)}} \right)} \right)}^H}\left( {{\mathbf{W}_{{k_\mathrm{D}}}} - \mathbf{W}_{{k_\mathrm{D}}}^{\left( m \right)}} \right)} \right\}}  \nonumber\\
			& + \mathrm{Tr}\left\{ {{{\left( {{\nabla _\mathbf{V}}{g}\left( {\mathbf{W}_{{k_\mathrm{D}}}^{\left( m \right)},{\mathbf{V}^{\left( m \right)}},p_{{k_\mathrm{U}}}^{\left( m \right)}} \right)} \right)}^H}\left( {\mathbf{V} - {\mathbf{V}^{\left( m \right)}}} \right)} \right\} + \sum\limits_{{k_\mathrm{U}} \in {\mathcal{K}_\mathrm{U}}} { {{{ {{\nabla _{{p_{{k_\mathrm{U}}}}}}{g}\left( {\mathbf{W}_{{k_\mathrm{D}}}^{\left( m \right)},{\mathbf{V}^{\left( m \right)}},p_{{k_\mathrm{U}}}^{\left( m \right)}} \right)} }}\left( {{p_{{k_\mathrm{U}}}} - p_{{k_\mathrm{U}}}^{\left( m \right)}} \right)} } \nonumber\\
			& \triangleq \widetilde {{g}}\left( {{\mathbf{W}_{{k_\mathrm{D}}}},\mathbf{V},{p_{{k_\mathrm{U}}}}\left| {\mathbf{W}_{{k_\mathrm{D}}}^{\left( m \right)},{\mathbf{V}^{\left( m \right)}},p_{{k_\mathrm{U}}}^{\left( m \right)}} \right.} \right) . 
	\end{align}}
\end{figure*}

Problem \eqref{max3} is also non-convex due to the objective function and the rank constraint C9. Note that $f$ and $g$ are concave functions, and thus the objective function in \eqref{max3} is a difference-of-concave function. Therefore, the SCA \cite{SCA} is applied to obtain a sub-optimal solution. Specifically, define the maximum number of iterations for SCA as $M$. In the $m$-th ($1 \le m \le M$) iteration, we construct a global overestimate of $g$ for a given feasible point $\left( {{\mathbf{W}_{k_\mathrm{D}}^{\left( m \right)}},{\mathbf{V}^{\left( m \right)}}},p_{k_\mathrm{U}}^{\left( m \right)} \right)$ by the first-order Taylor expansion, i.e., $\widetilde {g}\left( {{\mathbf{W}_{{k_\mathrm{D}}}},\mathbf{V},{p_{{k_\mathrm{U}}}}\left| {\mathbf{W}_{{k_\mathrm{D}}}^{\left( m \right)},{\mathbf{V}^{\left( m \right)}},p_{{k_\mathrm{U}}}^{\left( m \right)}} \right.} \right)$, which is given by \eqref{g_est} at the bottom of the page, where ${\nabla _{{\mathbf{W}_{{k_\mathrm{D}}}}}}{g}$, ${\nabla _\mathbf{V}}{g}$, and ${\nabla _{{p_{{k_\mathrm{U}}}}}}{g}$ denote the gradients of function $g$ with respect to ${\mathbf{W}_{{k_\mathrm{D}}}}$, $\mathbf{V}$, and ${p_{{k_\mathrm{U}}}}$, respectively.

Subsequently, for a given feasible point $\left( {{\mathbf{W}_{k_\mathrm{D}}^{\left( m \right)}},{\mathbf{V}^{\left( m \right)}}},p_{k_\mathrm{U}}^{\left( m \right)} \right)$ in the $m$-th iteration, a lower bound of the maximization problem in \eqref{max3} can be obtained by solving the following optimization problem.
\begin{align}\label{max4}
	& \mathop {\mathrm{maximize} } \limits_{{\mathbf{W}_{{k_\mathrm{D}}}}, \mathbf{V},{p_{{k_\mathrm{U}}}} } \quad \widetilde F \left( {\mathbf{W}_{{k_\mathrm{D}}}},\mathbf{V},{p_{{k_\mathrm{U}}}} \right) \\
	&\mathrm{s.t.} \quad \text{C3, C7, C8, C9}, \nonumber
\end{align}
where $\widetilde F\left( {\mathbf{W}_{{k_\mathrm{D}}}},\mathbf{V},{p_{{k_\mathrm{U}}}} \right)$ is defined as $\widetilde F\left( {\mathbf{W}_{{k_\mathrm{D}}}},\mathbf{V},{p_{{k_\mathrm{U}}}} \right) \triangleq  f-\widetilde {{g}}\left( {{\mathbf{W}_{{k_\mathrm{D}}}},\mathbf{V},{p_{{k_\mathrm{U}}}}\left| {\mathbf{W}_{{k_\mathrm{D}}}^{\left( m \right)},{\mathbf{V}^{\left( m \right)}},p_{{k_\mathrm{U}}}^{\left( m \right)}} \right.} \right)$. Note that the persistent non-convexity of problem\;\eqref{max4} stems from the rank-one constraint C9. Thus, the semidefinite relaxation (SDR) is adopted to relax constraint C9 by removing it. Then, the relaxed version of problem\;\eqref{max4} can be optimally solved with the aid of standard convex solvers such as CVX. Besides, the tightness of the rank relaxation is verified in the following theorem.
\newtheorem{theorem}{\textit Theorem}
\begin{theorem}
	If $P_{\max }^\mathrm{D} > 0$, the optimal beamforming matrices $\mathbf{W}_{k_\mathrm{D}}$ and $\mathbf{V}$, which satisfy $\mathrm{Rank} \left\lbrace  \mathbf{W}_{k_\mathrm{D}} \right\rbrace  \le 1$ and $\mathrm{Rank} \left\lbrace  \mathbf{V} \right\rbrace  \le 1$, can always be obtained.
\end{theorem}
\begin{IEEEproof}
Please refer to \cite[Appendix A]{SCA}.
\end{IEEEproof}

Then, the relaxed version of problem\;\eqref{max4} is iteratively solved until the increase of $\widetilde F\left( {\mathbf{W}_{{k_\mathrm{D}}}},\mathbf{V},{p_{{k_\mathrm{U}}}} \right)$ is less than the predefined convergence threshold ${\varepsilon _{{\mathrm{SCA}}}}$ or the maximum number of iterations $M$ is reached. Finally, the optimized UL powers $p_{k_\mathrm{U}}$ are outputted, and the optimized beamformers $\mathbf{w}_{k_\mathrm{D}}$ and $\mathbf{v}$ are obtained by performing the eigenvalue decomposition on ${\mathbf{W}_{{k_\mathrm{D}}}}$ and $\mathbf{V}$, respectively.
\subsection{Sub-Problem 3: Optimize $ {\mathbf{b}_{{k_\mathrm{U}}}} $ With Given $\left\{ {\tilde {\mathbf{t}},\tilde {\mathbf{r}}} \right\}$ and $\left\{ {{\mathbf{w}_{{k_\mathrm{D}}}},\mathbf{v},{p_{{k_\mathrm{U}}}}} \right\}$}
With given $\left\{ {\tilde {\mathbf{t}},\tilde {\mathbf{r}}} \right\}$ and $\left\{ {{\mathbf{w}_{{k_\mathrm{D}}}},\mathbf{v},{p_{{k_\mathrm{U}}}}} \right\}$, maximizing the receive SINR ${\gamma _{k_\mathrm{U}}^{\mathrm{U}}}$ of each UL user ${k_\mathrm{U}}$ with beamformer ${\mathbf{b}_{{k_\mathrm{U}}}}$ yields the maximization of the SSR. Specifically, let $\mathbf{w} \triangleq \sum\nolimits_{{k_\mathrm{D}} \in {\mathcal{K}_\mathrm{D}}} {{\mathbf{w}_{{k_{\mathrm{D}}}}}} \in {\mathbb{C}^{{N_\mathrm{t}} \times 1}}$ and $\mathbf{W} \triangleq \mathbf{ww}^H \in \mathbb{C}^{{N_\mathrm{t}} \times {N_\mathrm{t}}}$, we can obtain the optimal receive beamformer ${\mathbf{b}_{{k_\mathrm{U}}}}$ by solving the following optimization problem.
\begin{align}\label{max5}
	& \mathop {\mathrm{maximize} }\limits_{\mathbf{b}_{k_\mathrm{U}}} \quad \frac{\mathbf{b}_{k_\mathrm{U}}^H{\mathbf{h}_{\mathrm{UB},k_\mathrm{U}}}\left( {\tilde {\mathbf{r}}} \right)\mathbf{h}_{\mathrm{UB},{k_\mathrm{U}}}^H\left( {\tilde {\mathbf{r}}} \right){\mathbf{b}_{k_\mathrm{U}}}}{\mathbf{b}_{k_\mathrm{U}}^H{\mathbf{A}_{k_\mathrm{U}}}{\mathbf{b}_{k_\mathrm{U}}}} \\
	&\mathrm{s.t.} \quad \text{C1}, \nonumber
\end{align}
where ${\mathbf{A}_{k_\mathrm{U}}} \in \mathbb{C}^{{N_\mathrm{r}} \times {N_\mathrm{r}}}$ is defined as follows.
\begin{align}
	\mathbf{A}_{k_\mathrm{U}}
	& \triangleq \sum\limits_{i \in {\mathcal{K}_\mathrm{U}}\backslash \left\{ {k_\mathrm{U}} \right\}} {\mathbf{h}_{\mathrm{UB},i}\left( {\tilde {\mathbf{r}}} \right)\mathbf{h}_{\mathrm{UB},i}^H\left( {\tilde {\mathbf{r}}} \right){p_i}}  \nonumber\\
	& + \rho \left( {{\mathbf{H}_\mathrm{SI}}\left( {\tilde {\mathbf{t}},\tilde {\mathbf{r}}} \right)\left( {\mathbf{W} + \mathbf{V}} \right)\mathbf{H}_\mathrm{SI}^H\left( {\tilde {\mathbf{t}},\tilde {\mathbf{r}}} \right)} \right) + \sigma _\mathrm{U}^2{\mathbf{I}_{{N_\mathrm{r}}}} .
\end{align}
The optimal solution of problem\;\eqref{max5} is given by \cite{SCA}
\begin{equation}\label{receive_beam}
	\mathbf{b}_{k_\mathrm{U}}  = \frac{{{\mathbf{A}_{k_\mathrm{U}}^{ - 1}}{\mathbf{h}_{\mathrm{UB},{k_\mathrm{U}}}}\left( {\tilde {\mathbf{r}}} \right)}}{{{{\left\| {{\mathbf{A}_{k_\mathrm{U}}^{ - 1}}{\mathbf{h}_{\mathrm{UB},{k_\mathrm{U}}}}\left( {\tilde {\mathbf{r}}} \right)} \right\|}_2}}}.
\end{equation}

After obtaining the solutions of three sub-problems, the proposed AO algorithm iteratively solves the three sub-problems until the increase of $R_\mathrm{SSR}$ is less than the threshold $\varepsilon _{{\mathrm{AO}}}$ or the maximum number of iterations for AO is reached.
\subsection{Convergence and Complexity Analysis} \label{converge}
The convergence and computational complexity of the overall AO algorithm are analyzed as follows. Define the iteration index and maximum number of iterations for AO as $c$ and $C$, respectively, where $1 \le c\le C$. The convergence is ensured by the following inequality
\begin{align}
	& R_\mathrm{SSR} \left(  \tilde{\mathbf{t}}^{\left( c\right) },\tilde{\mathbf{r}}^{\left( c\right) },{\mathbf{w}_{{k_\mathrm{D}}}^{\left( c\right) }}, \mathbf{v}^{\left( c\right) },{p_{{k_\mathrm{U}}}^{\left( c\right) }}, {\mathbf{b}_{{k_\mathrm{U}}}^{\left( c\right) }} \right) \nonumber\\
	\mathop  \ge \limits^{\left( {{\alpha _5}} \right)} & R_\mathrm{SSR} \left(  \tilde{\mathbf{t}}^{\left( c\right) },\tilde{\mathbf{r}}^{\left( c\right) },{\mathbf{w}_{{k_\mathrm{D}}}^{\left( c\right) }}, \mathbf{v}^{\left( c\right) },{p_{{k_\mathrm{U}}}^{\left( c\right) }}, {\mathbf{b}_{{k_\mathrm{U}}}^{\left( c-1 \right) }} \right) \nonumber\\
	\mathop  \ge \limits^{\left( {{\alpha _6}} \right)} & R_\mathrm{SSR} \left(  \tilde{\mathbf{t}}^{\left( c\right) },\tilde{\mathbf{r}}^{\left( c\right) },{\mathbf{w}_{{k_\mathrm{D}}}^{\left( c-1\right) }}, \mathbf{v}^{\left( c-1\right) },{p_{{k_\mathrm{U}}}^{\left( c-1\right) }}, {\mathbf{b}_{{k_\mathrm{U}}}^{\left( c-1 \right) }} \right) \nonumber\\ 
	\mathop  \ge \limits^{\left( {{\alpha _7}} \right)} & R_\mathrm{SSR} \left(  \tilde{\mathbf{t}}^{\left( c-1\right) },\tilde{\mathbf{r}}^{\left( c-1\right) },{\mathbf{w}_{{k_\mathrm{D}}}^{\left( c-1\right) }}, \mathbf{v}^{\left( c-1\right) },{p_{{k_\mathrm{U}}}^{\left( c-1\right) }}, {\mathbf{b}_{{k_\mathrm{U}}}^{\left( c-1 \right) }} \right) . \nonumber\\  
\end{align}
The inequality marked by $\left( {{\alpha _5}} \right)$ holds because ${\mathbf{b}_{{k_\mathrm{U}}}^{\left( c\right) }}$ is the optimal receive beamformer for maximizing the SINR of each UL user. The inequality marked by $\left( {{\alpha _6}} \right)$ holds because ${\mathbf{w}_{{k_\mathrm{D}}}^{\left( c\right) }}$, $\mathbf{v}^{\left( c\right) }$, and ${p_{{k_\mathrm{U}}}^{\left( c\right) }}$ are the optimized transmit and AN beamformers and UL powers by the SCA. The inequality marked by $\left( {{\alpha _7}} \right)$ holds because $\tilde{\mathbf{t}}^{\left( c\right) }$ and $\tilde{\mathbf{r}}^{\left( c\right) }$ are the optimized positions of MAs searched by the proposed MVPSO. As a result, the SSR is non-decreasing during the iterations in the AO algorithm. Meanwhile, due to the finite communication resources, the SSR is always bounded. As such, the convergence of the overall algorithm is guaranteed. Moreover, the convergences are verified by the simulations in Section\;\ref{simu_conv}.

The computational complexity of the AO mainly arises from the search process of the MVPSO algorithm, the iterations of the AO and SCA algorithms, and the calculation of the receive beamformer. For the MVPSO in Algorithm\;\ref{MVPSO}, the complexities of calculating ${R_{\mathrm{SSR} }}\left( {\mathbf{u}_n^{\left( q \right)}} \right)$ and $\mathcal{P}\left( {\mathbf{u}_n^{\left( q \right)}} \right)$ in \eqref{fitness} are ${o_1} \triangleq \mathcal{O}\left( {N_\mathrm{t}}\left( {L_\mathrm{SI}^\mathrm{t}{N_\mathrm{r}} + \sum\nolimits_{{k_\mathrm{D}} \in {\mathcal{K}_\mathrm{D}}} {L_{\mathrm{BD},{k_\mathrm{D}}}^\mathrm{t}}  + \sum\nolimits_{{k_\mathrm{E}} \in {\mathcal{K}_\mathrm{E}}} {L_{\mathrm{BE},{k_\mathrm{E}}}^\mathrm{t}} } \right) \right.$ \\ $\left. + {N_\mathrm{r}}\left( {L_\mathrm{SI}^\mathrm{t} L_\mathrm{SI}^\mathrm{r} + \sum\nolimits_{{k_{\mathrm{U}}} \in {\mathcal{K}_\mathrm{U}}} {L_{\mathrm{UB},{k_\mathrm{U}}}^\mathrm{r}} } \right) \right)$ and ${o_2} \triangleq \mathcal{O} \left( { {N_\mathrm{t}} \choose 2 }+{ {N_\mathrm{r}} \choose 2 } \right)$, respectively. Thus, the total complexity of Algorithm\;\ref{MVPSO} is $\mathcal{O}\left( QN \bar I \left( o_1 + o_2 \right)  \right) $. For the SCA algorithm, let $M_\mathrm{SCA}$ denote the number of iterations, the complexity is $\mathcal{O}\left(M_\mathrm{SCA} \left(\left( K_\mathrm{D}+1 \right) N_\mathrm{t}^{3.5}+K_\mathrm{U}^{3.5}   \right)\right)  $ because of solving the SDR problem iteratively \cite{SCA}. For calculating the receive beamformer, the complexity is $\mathcal{O}\left( K_\mathrm{U} N_\mathrm{r}^3\right) $ due to the matrix inversion in \eqref{receive_beam}. Based on the above analyses, the computational complexity of the overall algorithm is $\mathcal{O}\left(C_\mathrm{AO} \left(  QN \bar I \left( o_1 + o_2 \right) + M_\mathrm{SCA} \left(\left( K_\mathrm{D}+1 \right) N_\mathrm{t}^{3.5}+K_\mathrm{U}^{3.5}   \right) \right.\right.$ \\ $\left.\left. + K_\mathrm{U} N_\mathrm{r}^3\right)  \right) $, where $C_\mathrm{AO}$ is the number of iterations for AO.
\subsection{Implementation Considerations of Proposed System} 
The above analyses focus on designing the resource allocation strategy for the proposed MA-aided secure FD multi-user system. In this sub-section, we discuss some practical considerations for the implementation of the proposed system.
\subsubsection{Antenna Movement}
The proposed system requires integrating an antenna positioning module with the conventional communication module at the FD BS. For large-scale systems, the motor-based MA system, which utilizes a step motor to facilitate the movement of the antenna along a sliding track, is particularly well-suited because the motor's driving power and the sliding track's length can be tailored to accommodate different antenna's weights and sizes \cite{MA1}. For small-scale systems, such as millimeter-wave (mmWave) and terahertz (THz) systems, micro-electromechanical systems (MEMS) can be employed to achieve low power consumption and high positioning accuracy \cite{MA2}.
\subsubsection{Adaptability of SI Cancellation Modules}
Unlike HD systems, the FD BS incorporates both analog and digital SI cancellation modules within its receive chains. These modules estimate the SI channel and then reconstruct the SI signal for cancellation. In conventional FPA FD systems, the SI channel remains relatively static due to fixed antenna positions. However, the MA-aided FD system faces dynamic changes in the SI channel caused by antenna movements, necessitating the efficient adaptability of the SI cancellation modules. A potential solution is the integration of machine learning algorithms, which can dynamically learn the characteristics of the varying SI channel, thus achieving excellent SI channel estimation \cite{SI_channel}.
\section{Simulation Results} \label{4}
In this section, we present the simulation results to evaluate the performance of the proposed scheme. 
\subsection{Simulation Setup and Benchmark Schemes}
\begin{table}[!t]
	\caption{Simulation Parameters}
	\label{tab1}
	\centering
	\begin{tabular}{|l|l|l|}
		\hline
		\multicolumn{1}{|c|}{\textbf{Parameter}} & \multicolumn{1}{c|}{\textbf{Description}} & \multicolumn{1}{c|}{\textbf{Value}} \\ \hline
		$\widetilde N$ & Number of antennas & 6 \\ \hline
		$A \times A$ & Moving region size  & $4\lambda \times 4\lambda$ \\ \hline
		$D$ & Minimum inter-MA distance & $\lambda/2$ \\ \hline
		$L$ & Number of channel paths  & $6$ \\ \hline
		$\rho$ & SI loss coefficient & -90dB   \\ \hline
		$\rho_0$ & Path loss at the reference distance & -40dB   \\ \hline
		$\alpha$ & Path loss exponent  & 2.8   \\ \hline
		$\sigma_\mathrm{U}^2$, $\sigma _{\mathrm{D}, k_\mathrm{D}}^2$, $\sigma _{\mathrm{E}, k_\mathrm{E}}^2$ & Average noise powers  & -90dBm   \\ \hline
		$P^\mathrm{D}_\mathrm{max}$ & Maximum DL transmit power  & 40dBm   \\ \hline
		$P^\mathrm{U}_\mathrm{max}$ & Maximum UL transmit power  & 10dBm  \\ \hline
		$K_\mathrm{D}$, $K_\mathrm{U}$, $K_\mathrm{E}$ & Numbers of users/Eves  & 4  \\ \hline
		$N$ & Number of particles  & 100   \\ \hline
		$Q$, $M$, $C$ & Maximum numbers of iterations  & 100   \\ \hline
		$\tau_\mathrm{t}$, $\tau_\mathrm{r}$ & Penalty factors  & 10   \\ \hline
		$c_1$, $c_2$ & Personal and global learning factors  & 1.4   \\ \hline
		$\varepsilon_\mathrm{SCA}$, $\varepsilon_\mathrm{AO}$ & Convergence thresholds  & $10^{-3}$   \\ \hline
	\end{tabular}
\end{table}
In the simulation, the UL users, DL users, and Eves are randomly and uniformly distributed in a cell centered on the FD BS with a radius of 600 meters (m). The FD BS is equipped with the same number of transmit and receive antennas, i.e., $N_\mathrm{t}=N_\mathrm{r}\triangleq \widetilde N$. We adopt the geometry channel model \cite{MA1,MA3}, where the numbers of all transmit and receive paths are identical, i.e., $L_\mathrm{SI}^\mathrm{t} = L_\mathrm{SI}^\mathrm{r} = L_{\mathrm{BD},k_\mathrm{D}}^\mathrm{t}= L_{\mathrm{BE},k_\mathrm{E}}^\mathrm{t}= L_{\mathrm{UB},k_\mathrm{U}}^\mathrm{r}\triangleq L$. In this way, the PRM $\mathbf{\Sigma}$ of the SI channel is a diagonal matrix, where each diagonal element follows the CSCG distribution $\mathcal{CN} \left( 0, \frac{\rho }{L} \right) $. For UL and DL channels, each element in PRVs follows the CSCG distribution $\mathcal{CN} \left( 0, \frac{\rho_0 d_k^{-\alpha} }{L} \right) $, where $\rho_0$ represents the path loss at the reference distance of 1 m, $\alpha$ is the path loss exponent, and $d_k$ denotes the propagation distance from the BS to DL user/Eve $k$ or from UL user $k$ to the BS. The AoDs and AoAs are assumed to be the independent and identically distributed random variables within the interval $\left[ 0, \pi \right] $. Unless specified otherwise, the default simulation parameters are set as shown in Table\;\ref{tab1}.
\begin{figure}[!t]
	\centering
	\subfloat[]{\label{PSO_conv}\includegraphics[width=0.5\columnwidth]{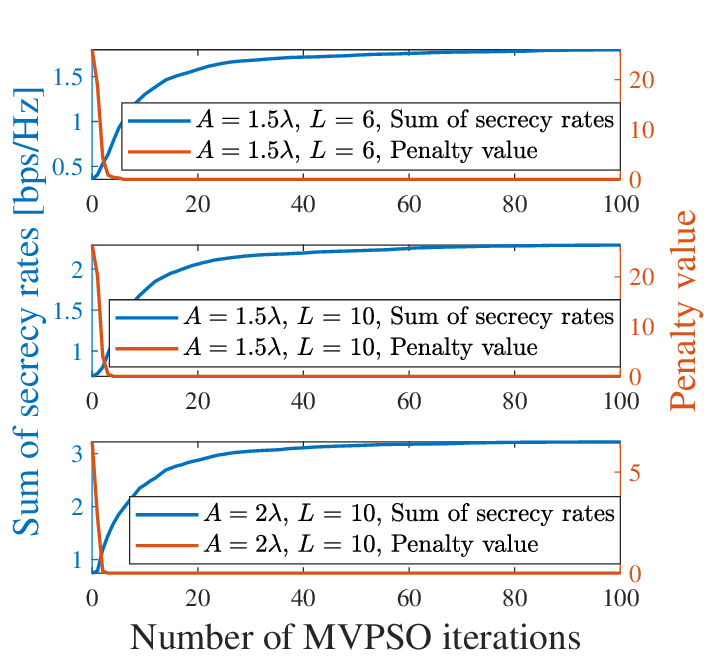}}
	\subfloat[]{\label{AO_conv}\includegraphics[width=0.5\columnwidth]{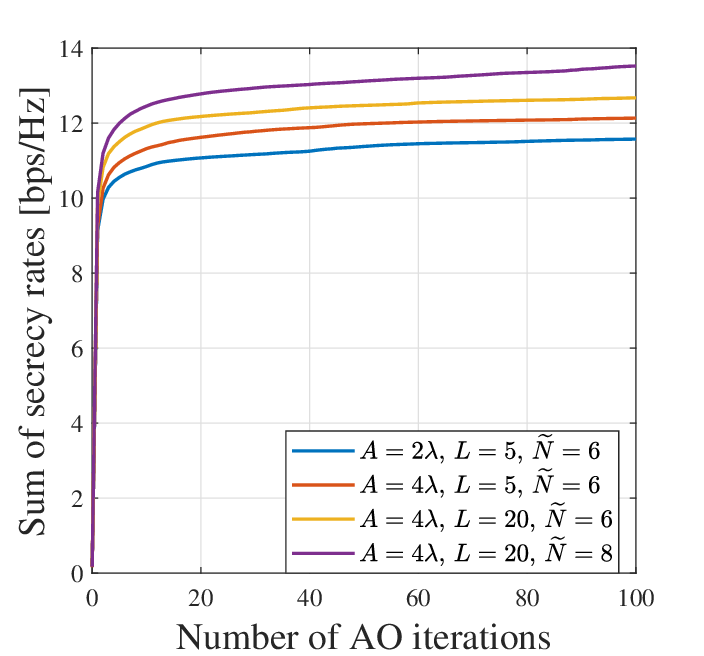}}
	\caption{Convergence evaluations of the proposed (a) MVPSO and (b) AO algorithms.}
	\label{conv}
\end{figure}

In the simulations, the proposed scheme is termed as ``Proposed''. For the MVPSO algorithm, we set $\bar I=3$ candidate velocities, which are derived from the weighted combinations of $\bar J=3$ velocity components. Specifically, the inertia weights for these three velocity components are respectively set as $\omega_1 = 0.9$, $\omega_2 = 0.75$, and $\omega_3 = 0.4$ to balance both the exploration and exploitation of the particles \cite{PSO}. Moreover, the combination weight vectors for all particles in \eqref{multi-v} are set as $\mathbf{c}_{n,1}=\left[ 1,0,0\right]^T $, $\mathbf{c}_{n,2}=\left[ 0,1,0\right]^T $, and $\mathbf{c}_{n,3}=\left[ 0,0,1\right]^T $. Besides, to fully demonstrate the advantages of the proposed scheme, we define the following state-of-the-art FPA-based and MA-based benchmark schemes.
\subsubsection{FPA-Based Schemes}
\begin{itemize}
	\item [$\circ$] 
	\textbf{FPA}: The BS is equipped with the transmit and receive FPA-based uniform
	planar arrays (UPAs) with $N_\mathrm{t}$ and $N_\mathrm{r}$ antennas, respectively. The UPAs and MA moving regions have the same aperture size.
	\item [$\circ$]
	\textbf{AS}: The BS is equipped with the transmit and receive FPA-based UPAs with $2N_\mathrm{t}$ and $2N_\mathrm{r}$ antennas, respectively. The UPAs and MA moving regions have the same aperture size. In a single AO iteration, $N_\mathrm{t}$ transmit antennas and $N_\mathrm{r}$ receive antennas are selected via exhaustive searches to maximize the SSR. 
\end{itemize}
\subsubsection{MA-Based Schemes With Various Antenna Position Optimization Algorithms}
\begin{itemize}
	\item [$\circ$]
	\textbf{Random position (RP)} \cite{MA3}: Randomly generate 100 pairs of $\tilde {\mathbf{t}}$ and $\tilde {\mathbf{r}}$ that satisfy constraints C4-C6. For each pair, solve sub-problems 2 and 3, and select the $\tilde {\mathbf{t}}$-$\tilde {\mathbf{r}}$ pair with the largest SSR.
	\item [$\circ$]
	\textbf{APO} \cite{MA4}: The transmit and receive regions are discretized into multiple grids of size $\frac{\lambda }{10} \times \frac{\lambda }{10}$. The MA positions are determined by the APO method, i.e., with other MAs fixed, the position of the current MA with the largest SSR, which meets constraints C5 and C6, is selected via exhaustive searches.
	\item [$\circ$]
	\textbf{PSO} \cite{Ding1}: This scheme optimizes the MA positions by the standard PSO algorithm.
\end{itemize}
\subsubsection{MA-Based Schemes With Various Transmission Strategies}
\begin{itemize}
	\item [$\circ$]
	\textbf{ZF}: This scheme employs the zero-forcing (ZF) receive beamformer instead of the optimal receive beamformer in \eqref{receive_beam} used in the proposed scheme.
	\item [$\circ$]
	\textbf{NoAN}: This scheme does not transmit the AN introduced in the proposed scheme. The transmit beamformers, $\mathbf{w}_{k_\mathrm{D}}$, are optimized with the same maximum DL power constraint, $P_{\max }^\mathrm{D}$.
	
	\item [$\circ$]
	\textbf{HD}: This scheme configures the BS to operate in the time-division HD mode. Thus, the SSR is penalized due to the half communication time compared to FD mode.
\end{itemize}
\subsection{Convergence Evaluations of Proposed Algorithms}\label{simu_conv}
In Fig.\;\ref{PSO_conv}, the convergence of the proposed MVPSO is evaluated with different moving region sizes and path numbers. As can be observed, the three SSRs increase with the number of iterations and tend towards stable values within 100 iterations, validating the convergence performance. Additionally, to verify the effectiveness of the proposed penalty function in \eqref{penalty_function}, we also illustrate the variations of the penalty values with the number of iterations. We can observe that the larger the moving region size, the fewer iterations are needed for the penalty value to reach zero. The three penalty values remain zero after 5 iterations, which ensures that the minimum inter-MA distance constraints C5 and C6 are satisfied.

Besides, the convergence evaluation of the overall AO algorithm is shown in Fig.\;\ref{AO_conv}. With different moving region sizes, numbers of paths, and numbers of MAs, the SSRs increase with the number of AO iterations and converge within 20 iterations, substantiating the previous discussions on the convergence of the AO algorithm in Section \ref{converge}.
\subsection{Channel Power Gains Under MVPSO}
\begin{figure*}[!t]
	\centering
	\subfloat[]{\label{channel_gain_BD}\includegraphics[width=0.66\columnwidth]{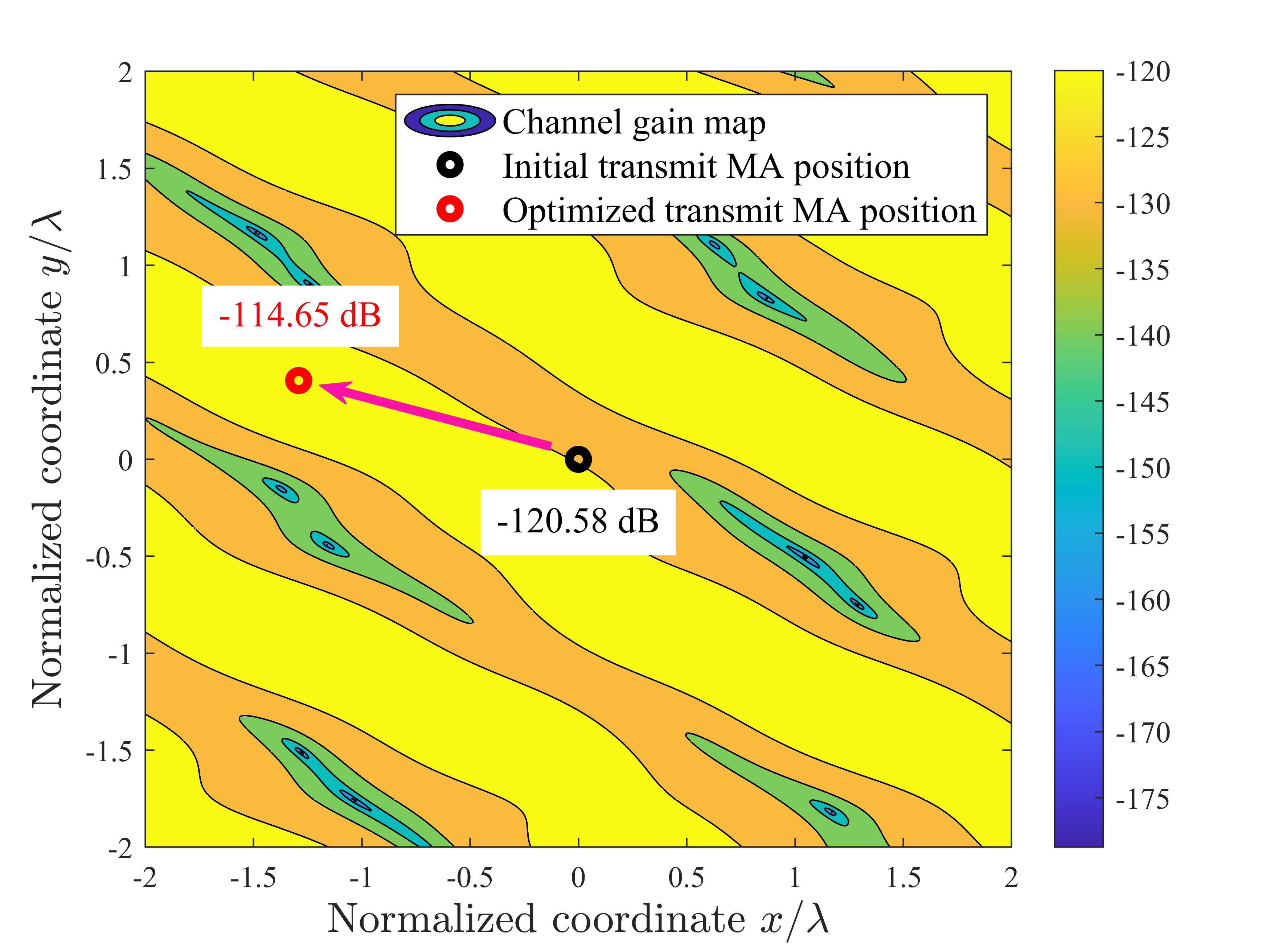}}
	\subfloat[]{\label{channel_gain_BE}\includegraphics[width=0.66\columnwidth]{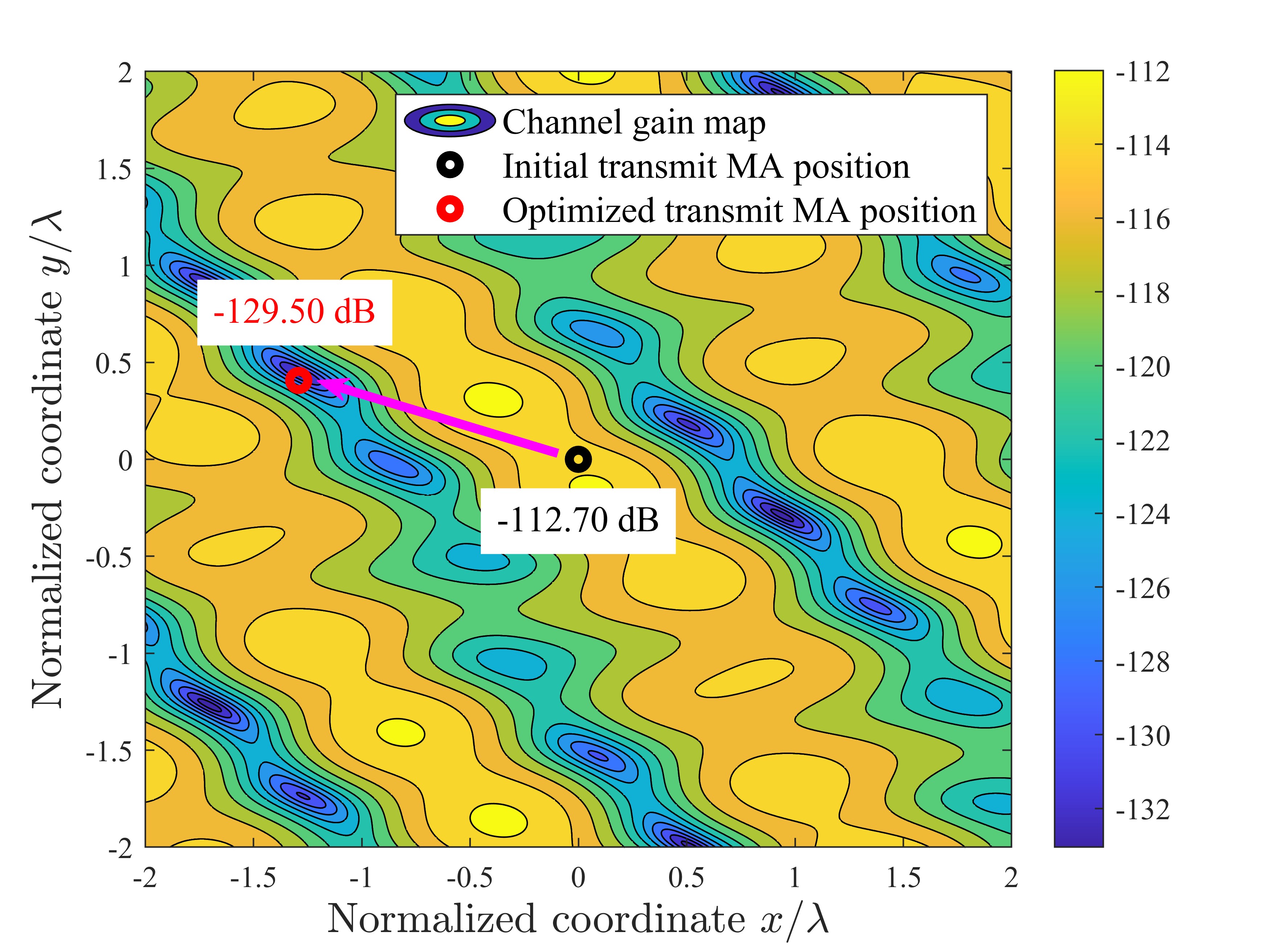}}
	\subfloat[]{\label{channel_gain_UB}\includegraphics[width=0.66\columnwidth]{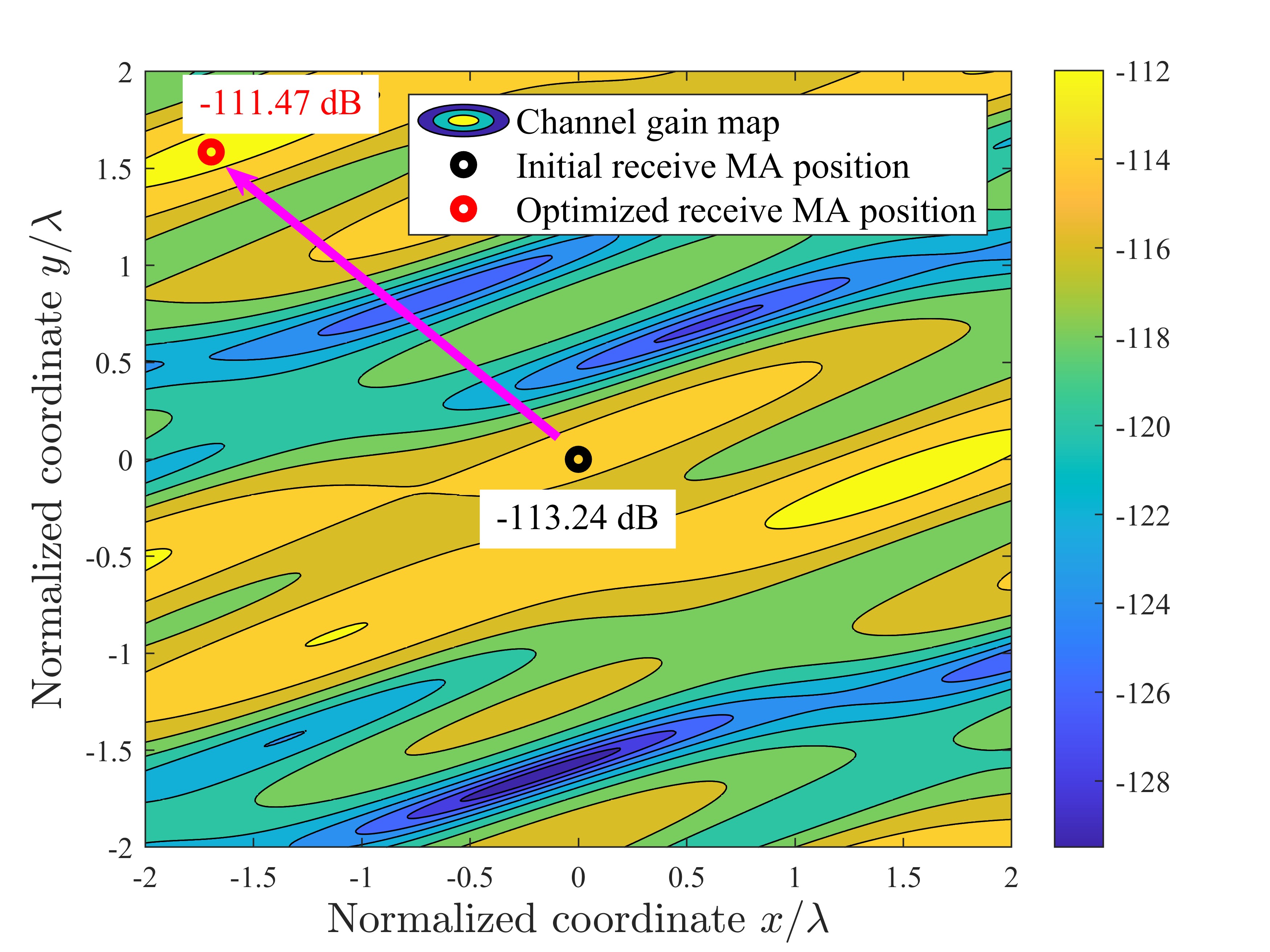}}\\
	\subfloat[]{\label{channel_gain_SI}\includegraphics[width=2\columnwidth]{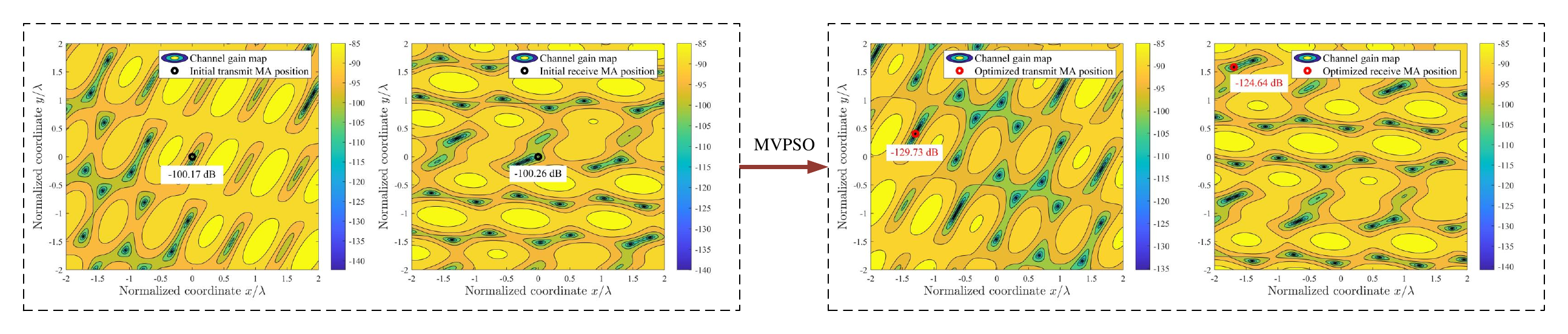}}
	\caption{Channel power gains (dB) (a) between the transmit MA of the BS and the single FPA of the DL user, (b) between the transmit MA of the BS and the single FPA of the Eve, (c) between the single FPA of the UL user and the receive MA of the BS, and (d) between the transmit MA and the receive MA of the BS. }
	\label{channel_gain}
\end{figure*}
\begin{figure}[!t]
	\centering
	\includegraphics[width=1\columnwidth]{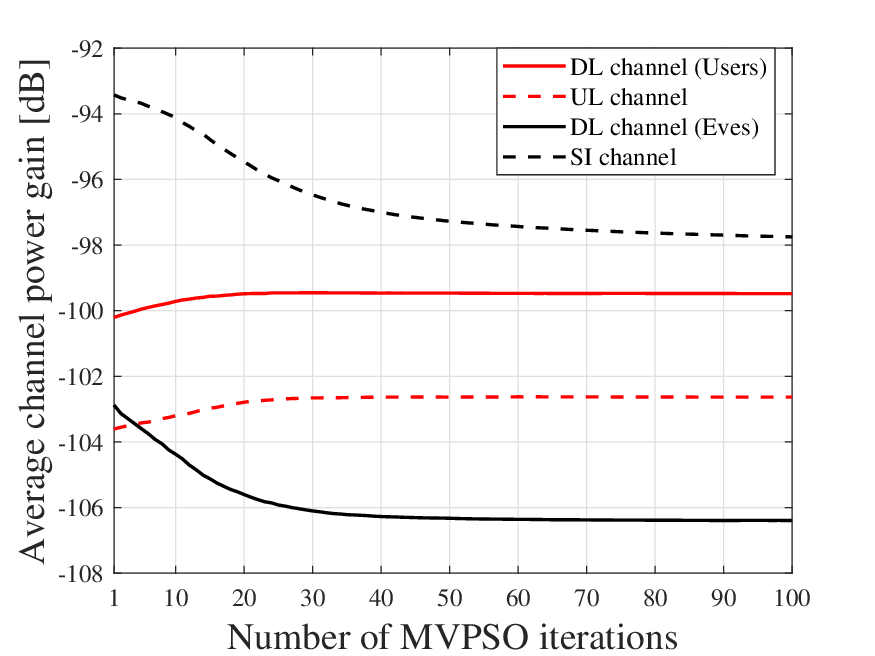}
	\caption{Average channel power gain for multiple users under MA position optimization using MVPSO.}
	\label{channel_power_mu}
\end{figure}
To further investigate the impact of antenna position optimization via the MVPSO on altering channel conditions, one realization of the channel power gains (in dB) versus the MA position is illustrated in Fig.\;\ref{channel_gain}. For ease of presentation, the numbers of transmit MA, receive MA, DL user, UL user, and Eve are set as 1, i.e., $\widetilde N=K_\mathrm{D}=K_\mathrm{U}=K_\mathrm{E}=1$. It is shown that due to the prominent small-scale fading in the spatial domain, for the DL and UL channels, each user or Eve has its unique channel gain map in the transmit or receive region (see Figs.\;\ref{channel_gain_BD}-\ref{channel_gain_UB}). For the SI channel, there are also corresponding channel gain maps in the transmit and receive regions before and after antenna position optimization (see Fig.\;\ref{channel_gain_SI}). As can be observed, for transmit MA, its position is initialized at the origin of the transmit region, with the channel power gains for the DL user, Eve, and the receive MA being -120.58 dB, -112.70 dB, and -100.17 dB, respectively. By employing the proposed MVPSO, the channel power gain of the optimized transmit MA's position for the DL user increases to -114.65 dB, while the channel power gains for the Eve and the receive MA decrease to -129.50 dB and -129.73 dB, respectively. Besides, for the receive MA, the optimized position, compared to the initial position, results in an increase of 1.77 dB in channel power gain for the UL user and a decrease of 24.38 dB in channel power gain for the transmit MA. These results indicate that even small movements of MAs can lead to significant changes in channel responses. Furthermore, antenna position optimization via the MVPSO comprehensively balances maximizing the channel power gains for the UL and DL users while minimizing those for the Eve and SI. These conclusions are also validated in Fig.\;\ref{channel_power_mu}, which illustrates the average channel power gain for the multi-user case under the MA position optimization by MVPSO. We can see that the channel power gains of DL and UL users increase, while those of SI and Eves decrease with the iterations, and the decreases are greater than the increases. This demonstrates that the position optimization of MAs through MVPSO can effectively reconfigure the channels based on multi-user system requirements, degrading performance-harmful channels (SI/Eves channels) while improving performance-beneficial ones (DL/UL users channels) to maximize the SSR.
\subsection{Performance Comparison With Benchmark Schemes}\label{per_section}
In this sub-section, we compare the performance of the proposed scheme with benchmark schemes.
\begin{figure}[!t]
	\centering
	\includegraphics[width=1\columnwidth]{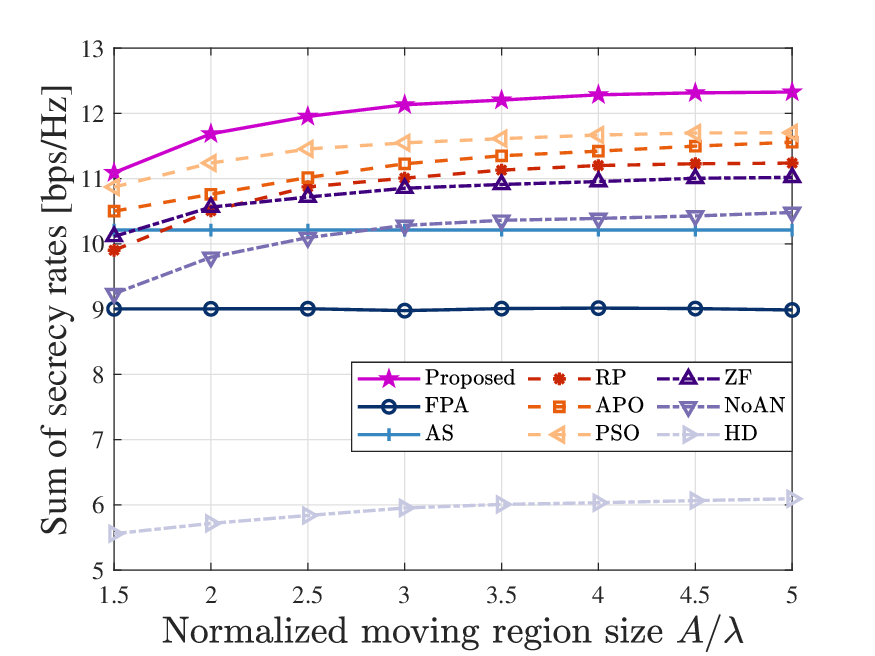}
	\caption{Sum of secrecy rates versus normalized moving region size.}
	\label{A}
\end{figure}
Fig.\;\ref{A} illustrates the SSR versus the normalized size of moving region for MAs. It can be observed that, the SSRs of all MA-based schemes increase with the normalized moving region size and gradually converge to stable values. This is because a larger moving region allows the MA to further explore spatial DoFs, but the limited number of paths restricts the diversity gains from increasing indefinitely. The fixed antenna configurations in the AS and FPA schemes render their performances unaffected by the moving region size, and the AS scheme can leverage spatial DoFs to some extent to improve the SSR compared to the FPA scheme. Furthermore, the proposed MVPSO outperforms the standard PSO, which in turn surpasses the conventional APO scheme. As the moving region size increases, the performance gap between the MVPSO and the PSO widens, while the gap between the PSO and the APO narrows. This is because the APO scheme maintains the fixed positions for other MAs while optimizing the current MA position, thereby disregarding the interdependencies among the MA positions. The standard PSO overcomes this limitation by optimizing all MA positions concurrently to prevent undesired sub-optimal solutions. However, its inability to balance exploitation and exploration hampers its effectiveness in discovering the optimal MA positions for the larger moving regions. Therefore, the MVPSO scheme stands out for its capability to simultaneously explore and exploit throughout the entire iteration process. In addition, the RP scheme performs the worst among all the considered antenna position optimization algorithms because it randomly initializes the MA positions without taking the actual channel conditions into account. Moreover, the results demonstrate that the AN and the FD mode effectively enhance the system's PLS performance. Unlike the HD systems, in FD mode, the transmission of AN can simultaneously safeguard both UL and DL users, and the reuse of spectrum further augments the SSR.
\begin{figure}[!t]
	\centering
	\includegraphics[width=1\columnwidth]{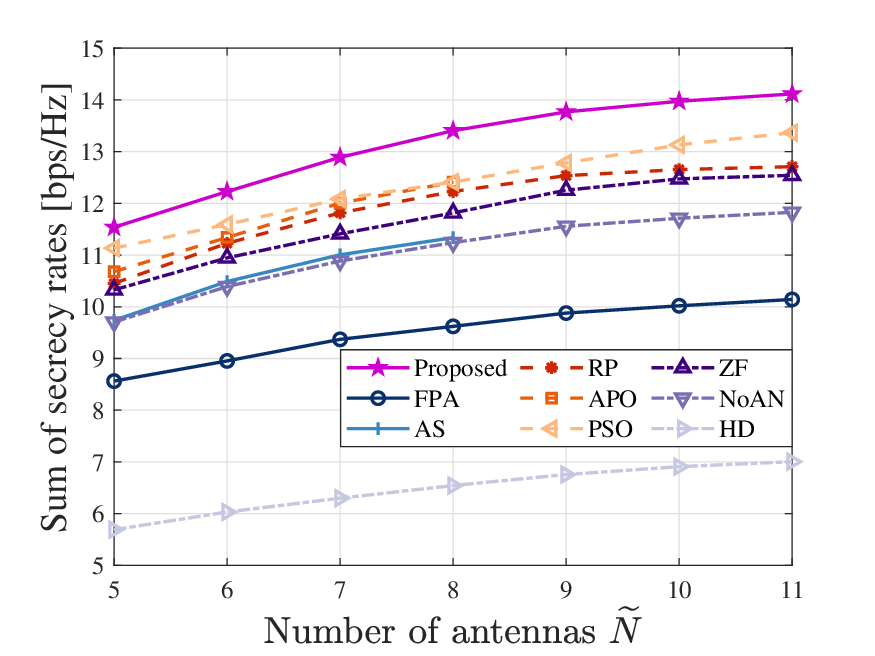}
	\caption{Sum of secrecy rates versus number of antennas.}
	\label{N}
\end{figure}

In Fig.\;\ref{N}, the SSRs of different schemes are compared versus the number of antennas. The APO and AS schemes are omitted when $\widetilde N > 8$ due to high computational complexities. For example, considering $\widetilde N = 10$ antennas in the $4\lambda \times 4\lambda$ transmit and receive regions, the APO scheme necessitates calculating $40 \times 40 \times 20 = 32000$ channel responses in a single AO iteration, and the number of total selections for the AS scheme is ${40 \choose 20} > 10^{11}$, which are hard to undertake. As the number of antennas increases, all schemes exhibit increases in the SSRs due to the enhanced spatial diversity gain and beamforming gain. We can find that the SSR growth in the FPA scheme is slower compared to the MA-based schemes. Specifically, the increase in the number of antennas from 5 to 11 brings the 18.44\% and 22.34\% increases in the SSRs for the FPA and the proposed MA-based scheme, respectively. This is because for the MA-based schemes, increasing the number of antennas allows for better exploitation of spatial DoFs by extensive movements within the designated regions. Consistently, the proposed scheme maintains optimal performance with varying numbers of antennas. Besides, we note that as the number of antennas increases, the APO scheme gradually approaches the SSR level nearly equivalent to that of the PSO scheme. This reveals that the standard PSO with a single velocity fails to fully search for the optimal solutions when the dimensions of the optimization variables, i.e., the number of antennas, are large, thus underscoring the necessity and effectiveness of the proposed MVPSO.

\begin{figure*}[!t]
	\centering
	\subfloat[]{\label{KU}\includegraphics[width=0.66\columnwidth]{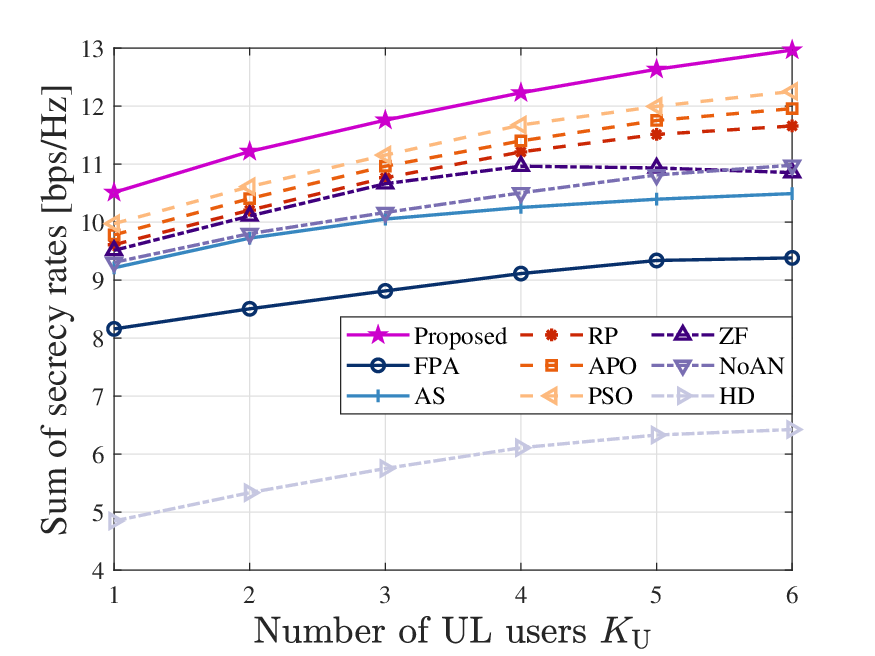}}
	\subfloat[]{\label{KD}\includegraphics[width=0.66\columnwidth]{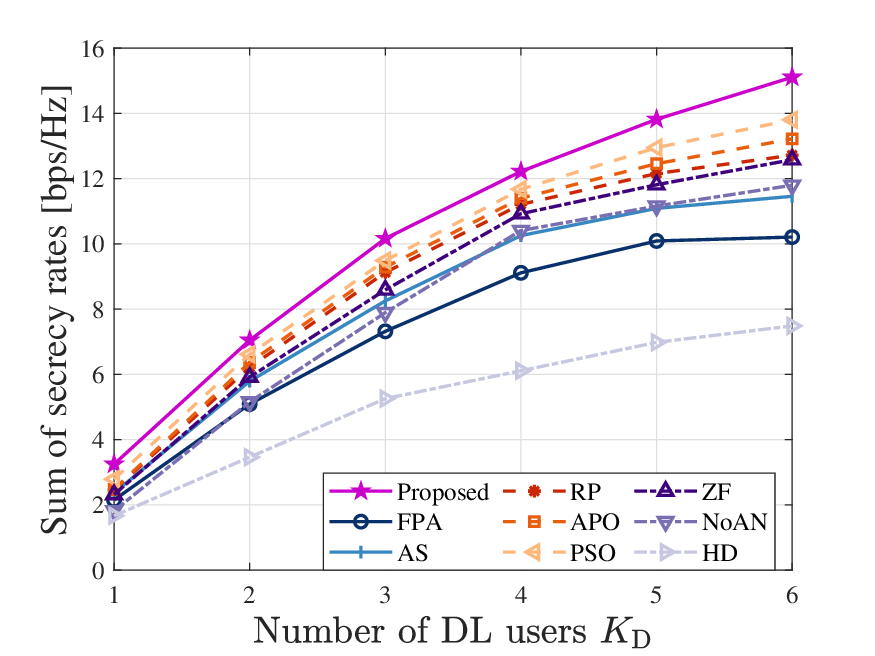}}
	\subfloat[]{\label{KE}\includegraphics[width=0.66\columnwidth]{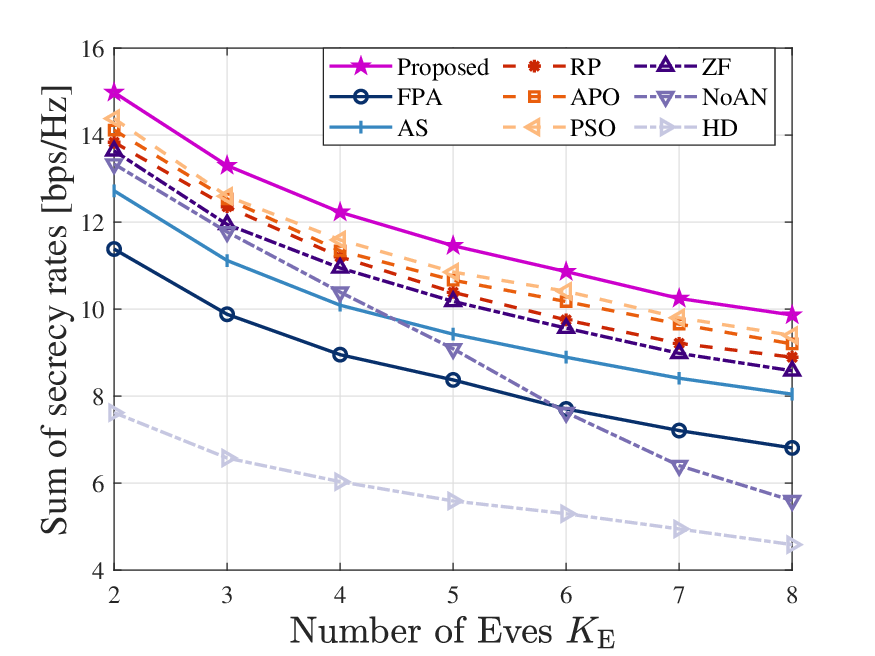}}
	\caption{Sums of secrecy rates versus numbers of UL users, (b) DL users, and (c) Eves.}
	\label{user_num}
\end{figure*}
To gain more insight, we present in Fig.\;\ref{user_num} the SSR for each scheme versus numbers of UL users, DL users, and Eves. As shown in Figs.\;\ref{KU} and \ref{KD}, the SSRs of the MA-based schemes (excluding the ZF scheme in Fig.\;\ref{KU}) consistently improve as the numbers of users increase, while the FPA-based schemes struggle when handling multiple users. This is because the flexible adjustment of the MA array can realize multi-beamforming with much less loss of the individual array gains in different directions. Besides, we note that the noise amplification caused by the ZF beamforming may deteriorate the SINRs of the UL users. Thus, the ZF scheme in Fig.\;\ref{KU} shows an SSR degradation with more UL users compared to the proposed optimal receive beamformer. Moreover, considering the worst-case scenario of the cooperative eavesdropping by multiple Eves who can completely cancel multi-user interference, the relationships between the performances of different systems and the number of Eves are depicted in Fig.\;\ref{KE}. Intuitively, the SSRs of all schemes decrease as the number of Eves increases. The reasons are as follows. With more Eves, the SINRs of cooperative UL and DL eavesdropping (as seen in the summation terms of \eqref{SINR_EU} and \eqref{SINR_ED}) increase. This coordinated effort undermines the systems' abilities to maintain secure communications, thereby diminishing the SSRs of all schemes. Fortunately, the proposed scheme still surpasses other benchmark schemes. However, the SSR of the NoAN scheme experiences a steep decline with an increasing number of Eves, approaching the performance of the HD scheme even at $K_\mathrm{E} = 8$. This demonstrates that due to the reuse of time-frequency resources in FD systems, the AN with effective beamforming can interfere with Eves' receptions while simultaneously protecting the UL and DL transmissions. Thus, AN is indispensable for secure FD transmission.

\begin{figure}[!t]
	\centering
	\subfloat[]{\label{rho_UL}\includegraphics[width=0.5\columnwidth]{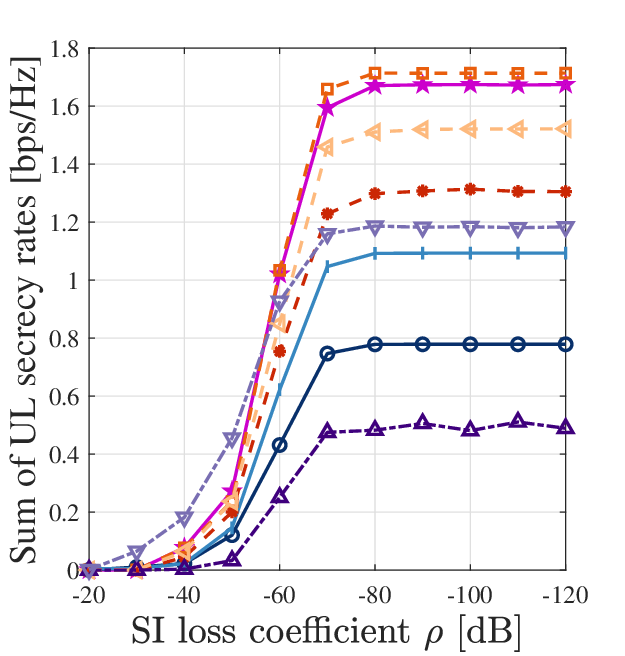}}
	\subfloat[]{\label{rho_DL}\includegraphics[width=0.5\columnwidth]{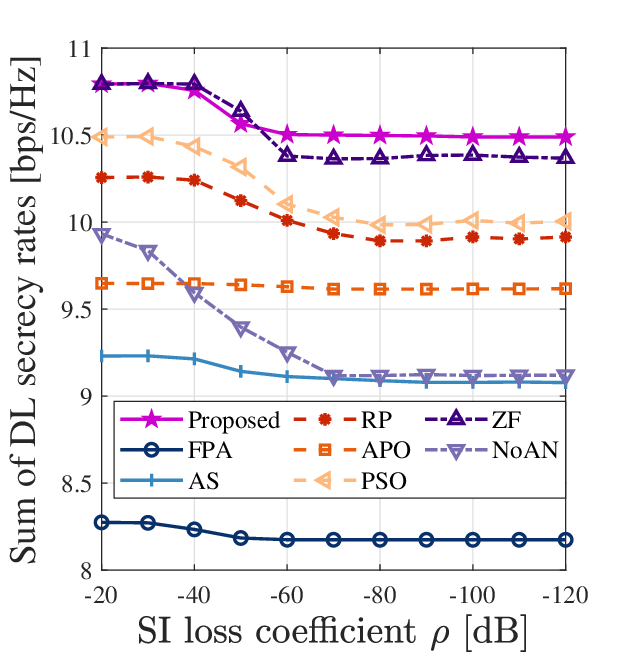}}
	\caption{Sums of (a) UL and (b) DL secrecy rates versus SI loss coefficient.}
	\label{rho}
\end{figure}
Subsequently, we investigate the impact of the SI loss coefficient on system performance. Figs.\;\ref{rho_UL} and \ref{rho_DL} respectively present the sums of UL and DL secrecy rates versus SI loss coefficient without the HD scheme. For the SI loss coefficient $\rho$, a smaller value in dB indicates a more powerful capability for SI cancellation. It can be observed that as $\rho$ decreases, the sums of UL and DL secrecy rates of all FD-based schemes increase and decrease, respectively. This is because, with the improved SI cancellation, the FD BS can reliably demodulate the UL users' information. Thus, the user scheduling policy allows for UL transmissions, resulting in an increased UL secrecy rate. However, for DL users, the undesired co-channel interference introduced by UL transmissions decreases their received SINRs, thus reducing the DL secrecy rate. Additionally, in case of weak SI cancellation, i.e., $-40\mathrm{dB}<\rho< -20\mathrm{dB}$, we observe the dominance of DL transmissions due to the severe SI affecting the reception of UL signals. Consequently, without UL transmissions, the performance of the ZF scheme closely mirrors that of the proposed scheme. On the contrary, when SI cancellation is robust, i.e., $\rho<-80\mathrm{dB}$, both UL and DL secrecy rates approach saturation. This is because the power of the residual SI becomes negligible compared to other interference perceived at the BS, rendering further improvements in SI cancellation ineffective in yielding significant gains. Besides, we notice that the UL secrecy rate of the APO scheme slightly surpasses that of the proposed scheme, but this improvement comes at the sacrifice of the DL secrecy rate. This indicates that, compared to the APO scheme, the proposed scheme more effectively leverages the DL transmission capability to maximize the sum of UL and DL secrecy rates. Moreover, the DL secrecy rate of the NoAN scheme in Fig.\;\ref{rho_DL} underperforms that of the APO scheme when $\rho < -40\mathrm{dB}$. This is because the discrete adaptation of the APO scheme to spatial channel variations results in its DL secrecy rate remaining relatively stable as the SI conditions change, a trend also seen in the FPA and AS schemes with discretely arranged antennas. In contrast, the continuous antenna movement in the NoAN scheme leads to a marked decline in secure DL communication performance as SI suppression improves. Furthermore, due to the absence of AN protection, the steady-state DL secrecy rate of the NoAN scheme is ultimately lower than that of the APO scheme.

\begin{figure*}[!t]
	\centering
	\subfloat[]{\label{PRM_err}\includegraphics[width=0.7\columnwidth]{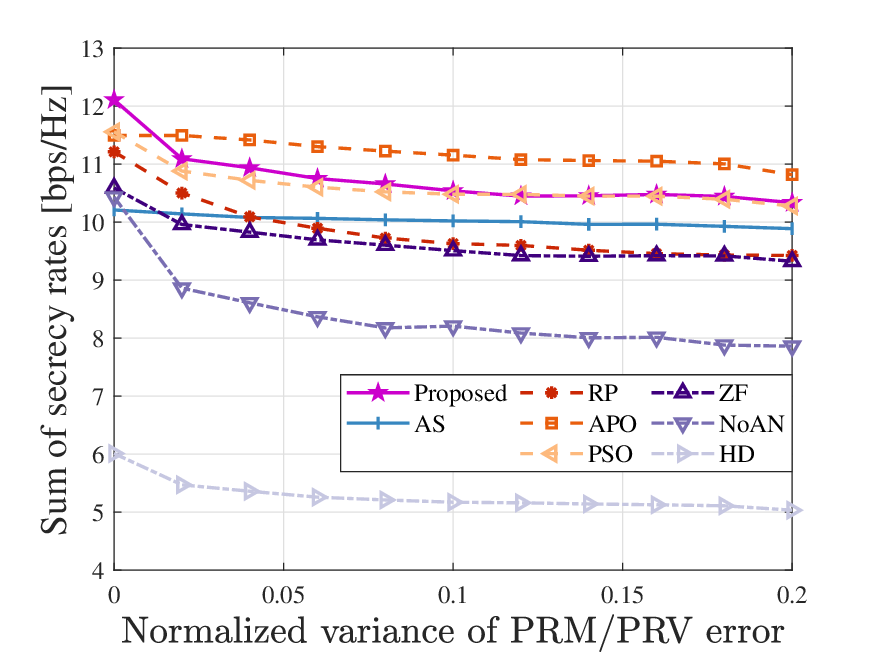}}
	\subfloat[]{\label{AoD_err}\includegraphics[width=0.7\columnwidth]{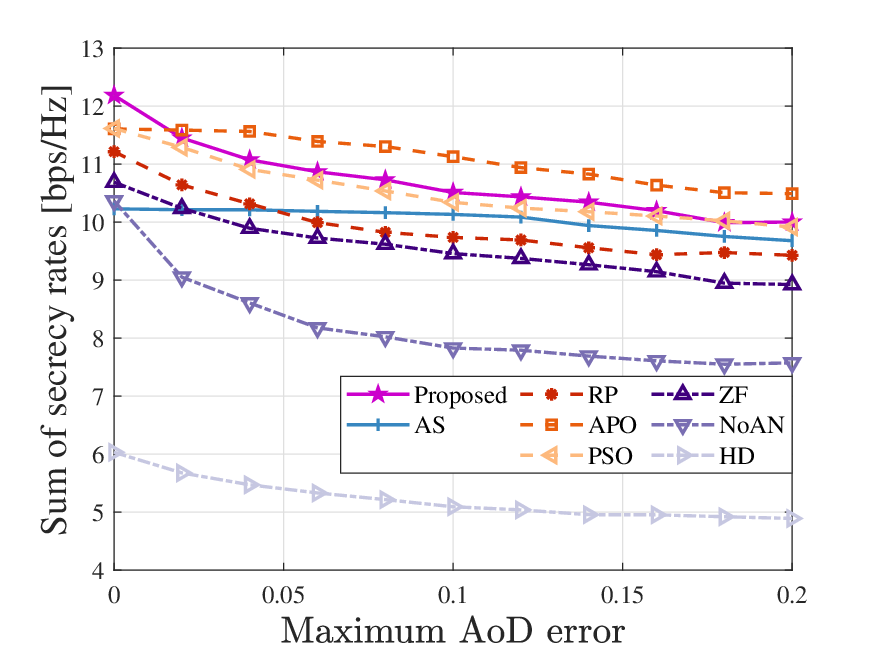}}
	\subfloat[]{\label{AoA_err}\includegraphics[width=0.7\columnwidth]{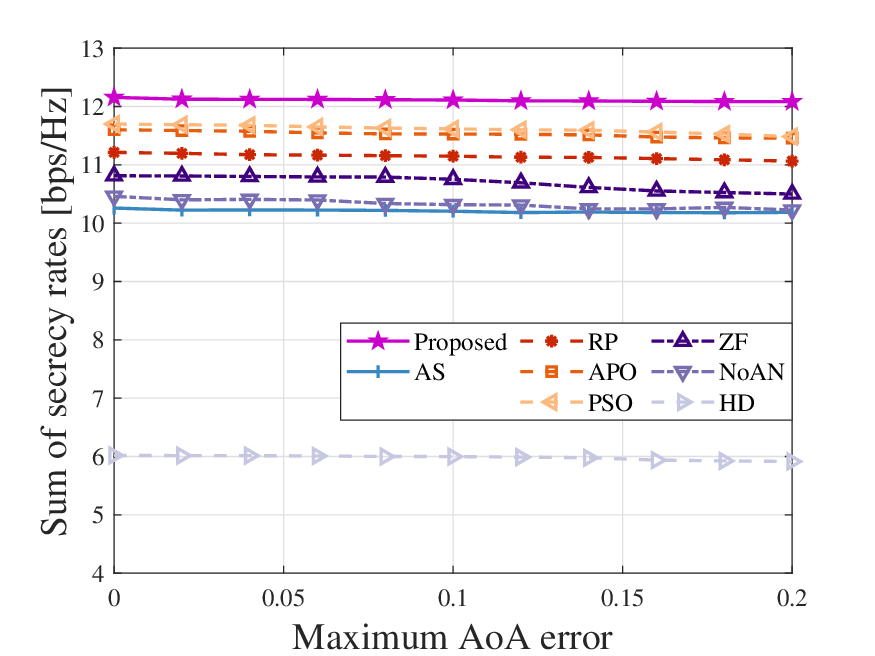}}
	\caption{Impact of the (a) PRM/PRV, (b) AoD, and (c) AoA errors on the performances of the proposed schemes.}
	\label{err}
\end{figure*}
In the above discussions, we assume that the BS possesses the perfect knowledge of the FRIs, i.e., the PRMs/PRVs of UL, DL, and SI channels, the AoDs of DL and SI channels, and the AoAs of UL and SI channels. Due to noise, limited training overhead, and multiple Eves, obtaining perfect FRIs is challenging in practice. Therefore, Fig.\;\ref{err} separately evaluates the impact of PRM/PRV, AoD, and AoA errors on system performances. The descriptions of the FRI errors reference \cite{MA4,MA5,MA11}. In each sub-figure, the remaining two types of FRIs are assumed to be perfectly known. To focus on the impact of imperfect FRIs on MA position optimization, the MA positions are determined by the estimated FRIs, while the calculations of the beamformers and UL powers are based on the actual channel responses \cite{MA5}. We neglect the FPA scheme because it does not involve antenna position optimization. From Fig.\;\ref{err}, it can be observed that the SSRs decrease with the normalized variance of PRM/PRV error and maximum AoD and AoA errors. Specifically, the proposed scheme experiences the performance losses of 14.65\%, 17.93\%, and 0.61\% due to PRM/PRV, AoD, and AoA errors, respectively. Compared to PRM/PRV and AoD errors, AoA error has a less detrimental effect on performances. The reasons are as follows. Because the AoA error only affects the calculations of UL and SI channels, the low UL powers make the deviations between the estimated and actual AoAs have an insignificant impact on UL transmissions. Thus, the MA position optimization based on the estimated AoA can still maintain excellent performance. Besides, we find that the APO scheme is less sensitive to PRM/PRV and AoD errors compared to the proposed scheme. This is because the APO scheme discretizes the moving regions into multiple grids, where the center of each grid serves as a candidate position. This discretization restricts MAs to discerning channel variations solely between different grids. In addition, the small PRM/PRV and AoD errors do not significantly impact the relative channel gains between different grids. Ultimately, the PRM/PRV and AoD errors have a minor impact on the APO scheme, yet the discretized movements fail to fully utilize spatial DoFs. Hence, in practical applications, we can strike a balance between fully exploiting spatial DoFs and mitigating the sensitivity to imperfect FRIs based on the SINR conditions. Specifically, in high SINR scenarios, the system does not need to rely heavily on MA movement to achieve higher spatial diversity gains, as the communication requirements can be easily met. Therefore, a discrete antenna positioning strategy \cite{MA6} can be adopted to enhance robustness against imperfect FRIs. Conversely, in low SINR scenarios, continuous MA movement is required to fully exploit the spatial DoFs and ensure reliable communications. In such cases, efficient FRI estimation techniques \cite{MA14,MA15} are essential to minimize system performance losses.

\section{Conclusion}\label{5}
In this paper, we proposed a new paradigm for secure FD multi-user systems, where an FD BS equipped with MAs simultaneously serves multiple UL and DL users while protecting their private information against multiple cooperative Eves. We formulated an optimization problem to maximize the SSR by jointly optimizing the MA positions, the transmit, receive, and AN beamformers at the BS, and the UL powers. To tackle this non-convex problem, an AO algorithm was proposed to decompose the original problem into three sub-problems and iteratively solve them. Specifically, the MA positions are updated by the proposed MVPSO, which applies multiple candidate velocities instead of the single velocity used in the standard PSO algorithm to determine the superior MA positions. Additionally, the SSR is reformulated as a difference-of-concave function to optimize the transmit/AN beamformers and the UL powers via the SCA algorithm. Moreover, the optimal receive beamformer is derived as a closed-form solution. Simulation results revealed the effectiveness of the proposed MVPSO in determining the MA positions compared to the standard PSO and other state-of-the-art antenna position optimization algorithms. Besides, the results showed that the additional DoFs in antenna movement provided by MAs benefit the enhancement of signal power, the mitigation of various interferences, and the flexibility of multi-beamforming in the proposed system, thereby leading to excellent security performance. In addition, the results demonstrated that the AN transmitted by the FD BS can protect both UL and DL users simultaneously, a capability that cannot be achieved with an HD BS. Furthermore, we evaluated the impact of the discrepancies between the estimated and actual FRIs on system performance, which provides valuable references for the applications of the proposed scheme in practice.

\newpage

\vfill

\end{document}